\documentclass[acmsmall]{acmart}

\usepackage{booktabs} 
\usepackage[table]{xcolor} 
\usepackage{graphicx}
\usepackage{caption}
\usepackage{amsmath}
\usepackage{geometry}
\usepackage{float}
\usepackage{threeparttable}
\usepackage{tikz}
\tikzset{arr/.style={-{Latex}}} 
\usetikzlibrary{arrows.meta,positioning,fit,calc,shapes.misc}

\usepackage{pgfplots}
\usepackage{adjustbox}
\pgfplotsset{compat=1.18}
\usepackage{url}
\usepackage{tcolorbox}
\usepackage{amsmath}
\usepackage{pifont}
\usepackage{xcolor}
\usepackage{multirow} 
\usepackage[table]{xcolor}
\usepackage{subcaption} 
\usepackage{tcolorbox}         
\usepackage{adjustbox}  
\usepackage[ruled,vlined,linesnumbered]{algorithm2e}
\usepackage{wrapfig}

\newcommand{\toolns}{\textit{Chameleon}\xspace}
\newcommand{\tool}{\textit{Chameleon}\xspace}

\AtBeginDocument{%
  }

\setcopyright{cc}
\setcctype{by}
\acmDOI{10.1145/3832148}
\acmYear{2026}
\acmJournal{PACMSE}
\acmVolume{3}
\acmNumber{ISSTA}
\acmArticle{ISSTA057}
\acmMonth{10}
\acmSubmissionID{issta26main-p473-p}
\received{2026-01-30}
\received[accepted]{2026-04-16}

\begin{document}

\title{Environmental Injection Attacks against GUI Agents in Realistic Dynamic Environments}

\author{Yitong Zhang}
\orcid{0009-0000-1138-4503}
\affiliation{%
  \institution{Tsinghua University}
  \department{College of AI}
  \city{Beijing}
  \country{China}
}
\email{22373337@buaa.edu.cn}

\author{Ximo Li}
\orcid{0009-0008-6594-9897}
\affiliation{%
  \institution{Tsinghua University}
  \department{Department of Computer Science and Technology}
  \city{Beijing}
  \country{China}
}
\email{lixm23@mails.tsinghua.edu.cn}

\author{Liyi Cai}
\orcid{0009-0002-7848-6007}
\affiliation{%
  \institution{Peking University}
  \department{School of Computer Science}
  \city{Beijing}
  \country{China}
}
\email{cailiyi@stu.pku.edu.cn}

\author{Jia Li}
\orcid{0000-0002-5579-8852}
\affiliation{%
  \institution{Tsinghua University}
  \department{College of AI}
  \city{Beijing}
  \country{China}
}
\email{jia\_li@mail.tsinghua.edu.cn}

\renewcommand{\shortauthors}{Yitong Zhang, Ximo Li, Liyi Cai and Jia Li}

\begin{abstract}
Graphical User Interface (GUI) agents are increasingly deployed to interact with online web services, yet their exposure to open-world content renders them vulnerable to Environmental Injection Attacks (EIAs). In these attacks, an attacker can inject crafted triggers into a website to manipulate the behavior of other users’ GUI agents.
In this paper, we find that most existing EIA studies fall short of realism. In particular, they fail to capture the dynamic nature of real-world websites, often assuming that a trigger’s on-screen position and surrounding visual context remain largely consistent between training and testing.
To better reflect practice, we introduce a realistic \textit{dynamic-environment threat model} in which the attacker is a regular user and the trigger is embedded within a dynamically changing environment. Under this threat model, existing approaches largely fail, suggesting that their effectiveness in exposing GUI agent vulnerabilities has been overestimated.

To expose the hidden vulnerabilities of existing GUI agents effectively, we propose \toolns, an attack framework with two key components designed for dynamic environments.
\ding{182} To synthesize more realistic training data, we introduce \textit{LLM-Driven Environment Simulation}, which automatically generates diverse, high-fidelity webpage simulations that mimic the variability of real-world dynamic environments.
\ding{183} To optimize the trigger more effectively, we introduce \textit{Attention Black Hole}, which converts attention weights into explicit supervisory signals.
We evaluate \tool on six realistic websites and four representative LVLM-powered GUI agents. Across these settings, it significantly outperforms existing methods.
Ablation studies confirm that both components are critical to performance, and a closed-loop sandbox experiment further demonstrates that \tool can successfully hijack agent behavior in conditions that closely mirror real-world usage.
Our results uncover a critical, previously underexplored vulnerability of GUI agents in realistic dynamic environments and establish a robust foundation for future research on defenses for open-world GUI agent systems.
\end{abstract}




\maketitle

\section{Introduction}
\label{sec:introduction}

With the rapid development of Large Language Models (LLMs) and Large Vision-Language Models (LVLMs)~\cite{gpt4o, wang2024qwen2, bai2025qwen2, wu2024deepseek, nong2024mobileflow}, a new class of powerful agents called GUI agents has emerged to interact with Graphical User Interfaces (GUIs)~\cite{hong2024cogagent, qin2025ui, wu2024atlas, zhang2024large}.
These GUI agents can execute increasingly sophisticated tasks, moving beyond simple dialogue to perform complex operations like website manipulation~\cite{zhou2023webarena}. 
A key characteristic of these agents is their ability to autonomously access and act upon live internet content. While this capability greatly expands their functionality, it simultaneously introduces novel security risks inherent to open-world interaction~\cite{nguyen2024gui,wang2024gui, johnson2025manipulating, yang2025riosworld, yang2025mla, ye2025visualtrap, chen2025obvious}.

A growing body of work has shown that GUI agents are particularly vulnerable to Environmental Injection Attacks (EIAs), which originate from the external environment rather than from the users themselves~\cite{liao2024eia,ma2024caution, aichberger2025attacking,zhang2024attacking}.
Recently, EIA research has increasingly adopted a practical assumption that the attacker is a malicious regular user without administrative privileges and can only influence the webpage through normal content uploads.
For instance, on an e-commerce platform, a malicious attacker could upload a specially crafted trigger image disguised as a product photo. When another user's GUI agent encounters this trigger while browsing the website, it might be automatically induced to navigate to a specific promotional site, all without the user's explicit instruction~\cite{zhao2025robustness, lu2025eva}.
 
\begin{figure}[t]
  \centering
  \subcaptionbox{}[0.48\linewidth]{%
    \includegraphics[width=\linewidth]{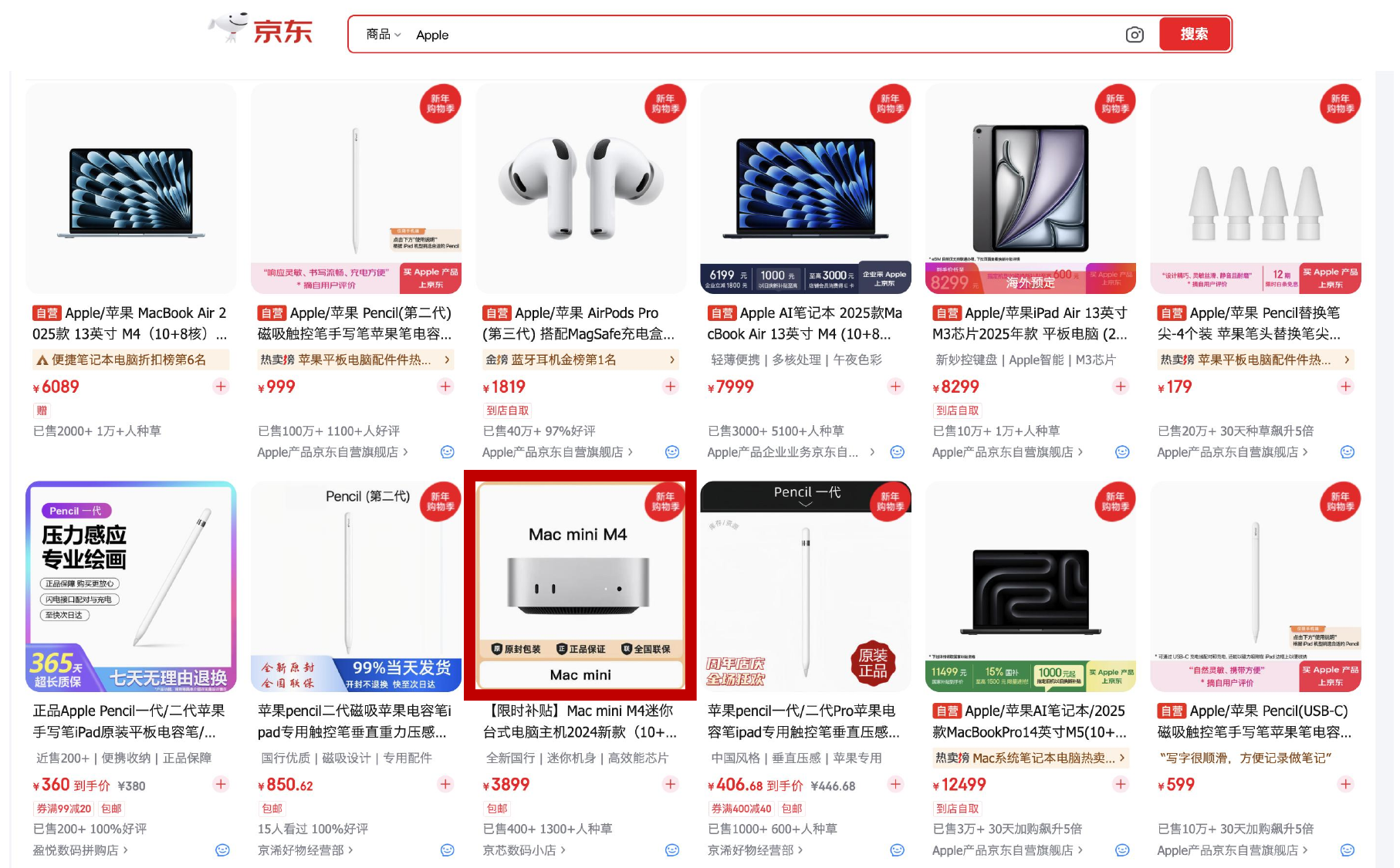}}
  \hspace{0.02\linewidth}
  \subcaptionbox{}[0.48\linewidth]{%
    \includegraphics[width=\linewidth]{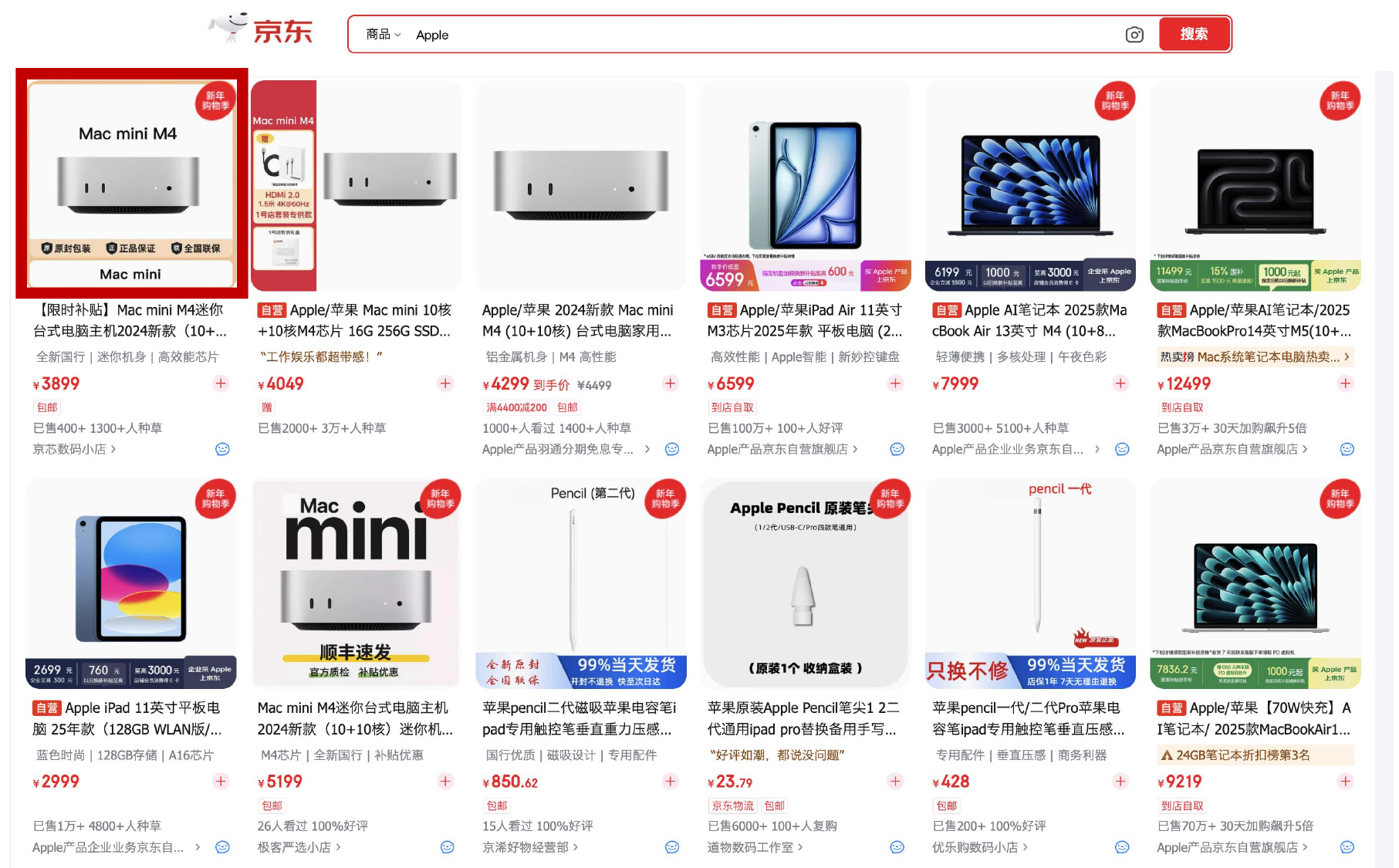}}
    \vspace{-0.1in}
    \caption{
    An illustration of the \textit{dynamic environment} in realistic GUI agent application scenarios. The two screenshots show consecutive searches for ``Apple'' on the same \href{https://search.jd.com/}{e-commerce platform}. We assume the image highlighted by the red box is a trigger uploaded by an attacker.
    }
  \label{fig:motivation}
  \vspace{-0.14in}
\end{figure}

However, we find that \textbf{most existing EIAs still fall short of realism}. In particular, prior work~\cite{wu2024dissecting, aichberger2025attacking, zhao2025robustness} often evaluates triggers under static or only slightly varying environments, implicitly assuming that the trigger’s \textit{\textbf{on-screen position}} and \textit{\textbf{surrounding visual context}} remain largely consistent between training and testing. This assumption conflicts with how modern websites operate. In practice, layouts and nearby content change continuously due to dynamic ranking and recommendation updates, advertisement placement, and frequent content refresh.
To illustrate this gap, consider a common e-commerce scenario. Figure~\ref{fig:motivation} shows two screenshots from JD.com\footnote{\url{https://search.jd.com/}}, collected in two separate sessions using the same query (``Apple''). Even in this simple setting, the trigger image highlighted in red appears at different positions on the page, and its surrounding context, including nearby products, banners, and text snippets, also changes substantially. Such environmental dynamics pose a fundamental challenge to existing attacks. Many EIA methods optimize and evaluate trigger images in static or slightly varying environments. When deployed in realistic dynamic environments, shifts in trigger position and changes in surrounding visual context may sharply reduce attack effectiveness. These observations motivate a novel threat model that explicitly accounts for environmental dynamism, rather than relying on overly idealistic static assumptions.

% preamble:
% \usepackage{graphicx}
% \usepackage{subcaption}
\captionsetup[subfigure]{justification=centering,font=small,skip=2pt}

In this paper, we formally define a realistic \textbf{dynamic-environment threat model}, in which the attacker can hijack the GUI agent by uploading trigger images as a regular user. Crucially, these triggers are embedded within a dynamically changing environment, where their positions and surrounding visual contexts may continuously change and are largely beyond the attacker’s control.
Under this threat model, our experiments in Section~\ref{sec:rq1} show that trigger images optimized using existing environmental injection attack approaches exhibit near-zero attack success rates~\cite{aichberger2025attacking, wu2024dissecting, zhao2025robustness, madry2017towards}. 
This is largely because prior methods optimize triggers under static or slightly varying environments, and thus fail to generalize to dynamic environments. Such a mismatch can mislead security analysis and hinder the development of effective defenses for GUI agents in realistic deployments.

\begin{figure}[t]
  \centering
  \subcaptionbox{Successful case}[0.48\linewidth]{%
    \includegraphics[width=\linewidth]{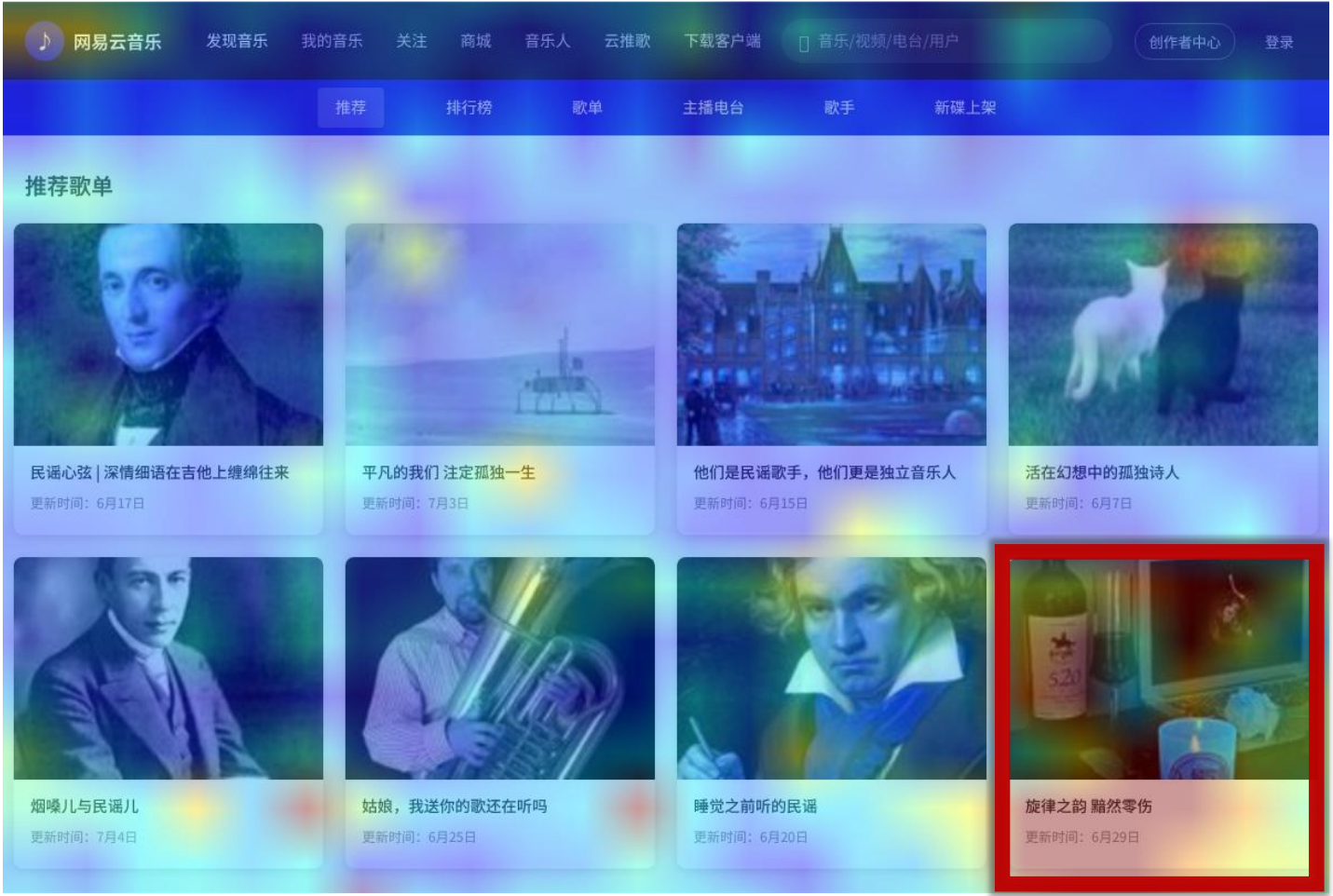}}
  \hspace{0.02\linewidth}
  \subcaptionbox{Unsuccessful case}[0.48\linewidth]{%
    \includegraphics[width=\linewidth]{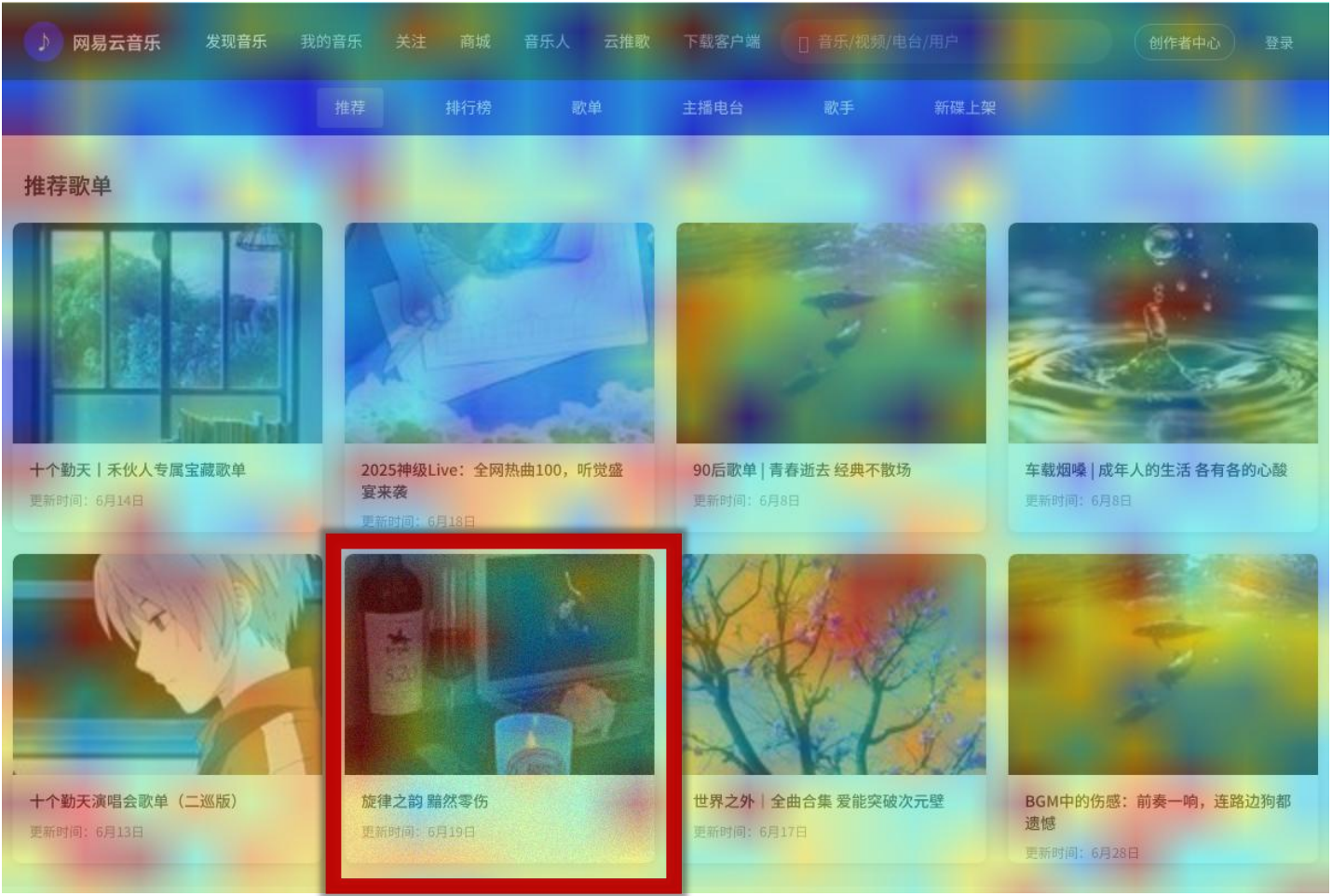}}
    % \vspace{-0.1in}
    \caption{Attention maps for the two cases. The red box marks the trigger image; warmer colors indicate higher attention. In the successful case, attention is concentrated on the trigger image region, whereas in the unsuccessful case, attention is dispersed across the screenshot.}
  \label{fig:attn-cases}
  % \vspace{-0.14in}
\end{figure}

To more fully expose the vulnerabilities of existing GUI agents in realistic deployments, we propose \toolns. This novel attack framework is specifically designed to ensure that trigger images remain effective within dynamic environments, where on-screen positions and surrounding visual contexts vary unpredictably.
\tool addresses two key challenges.
\ding{182} \textbf{How to synthesize large-scale training data that captures dynamic environments?}
Achieving robust attacks in dynamic environments requires a large number of realistic training samples that cover diverse trigger placements and contextual variations. However, manually collecting webpage samples with diverse layouts and surrounding content is laborious and time-consuming~\cite{koh2024visualwebarena, liu2025hijacking, kasuga2024cxsimulator}.
To overcome this challenge, we introduce \textit{\textbf{LLM-Driven Environment Simulation}}. Leveraging the generative capabilities of LLMs, we automatically construct high-fidelity simulations of target websites and systematically vary both the trigger placement and its surrounding context. This process produces a large volume of realistic screenshots, enabling the optimized trigger to generalize effectively across dynamic environments.
\ding{183} \textbf{How to train the trigger more effectively?}
We find that even with sufficient realistic training data, attack performance can remain suboptimal.
Our preliminary observations suggest that the agent’s attention is often distracted by constantly changing environments. As shown in Figure~\ref{fig:attn-cases}, successful attacks occur when attention remains concentrated on the trigger, whereas unsuccessful cases arise when attention is dispersed across other salient interface elements.
To address this issue, we propose \textit{\textbf{Attention Black Hole}}, which enables more effective trigger optimization in dynamic environments. The key idea is to explicitly guide the model to focus on the trigger region while suppressing interference from non-trigger areas. Concretely, \textit{Attention Black Hole} introduces an explicit supervisory signal derived from attention weights~\cite{lu2025eva, zhang2024enhancing}, encouraging attention to consistently concentrate on the trigger image. This design improves attack reliability under dynamic visual distractions.

% preamble:
% \usepackage{graphicx}
% \usepackage{subcaption}
\captionsetup[subfigure]{justification=centering,font=small,skip=2pt}

% : attention concentrates on the trigger image region.
% : attention is dispersed over the screenshot.

We conduct an in-depth evaluation of our proposed \toolns. 
We first construct six highly realistic datasets that simulate widely used websites with dynamically varying images and text.
We assess four representative GUI agents (UI-TARS-7B-DPO~\cite{qin2025ui}, OS-Atlas-Base-7B~\cite{wu2024atlas}, Qwen2-VL-7B~\cite{wang2024qwen2}, and LLaVA-1.5-13B~\cite{liu2023visual}) and find that \tool substantially outperforms all baselines across all websites~\cite{madry2017towards, wu2024dissecting, aichberger2025attacking}. 
We further observe favorable cross-model generalization across similar models. 
An ablation study shows that both \textit{LLM-Driven Environment Simulation} and \textit{Attention Black Hole} are necessary components of \toolns.
Finally, we evaluate several practical defenses and find that many existing strategies are ineffective against \toolns without substantially harming user experience, highlighting the need for new defense mechanisms tailored to GUI agents.

The contributions of this paper are threefold.
\begin{itemize}
\item We formalize a realistic \textit{dynamic-environment threat model} for EIA. Under this novel threat model, we demonstrate that existing EIA approaches are largely ineffective.

\item We propose \toolns, a new environmental injection attack framework that is effective under the realistic dynamic-environment threat model, enabling the discovery of previously hidden vulnerabilities in existing GUI agents.

\item We conduct extensive experiments on four representative GUI agents across six websites. The results obtained through \tool uncover previously underexplored vulnerabilities in existing GUI agents, highlighting the urgent need for more robust security mechanisms.
\end{itemize}

\section{Background and Related Work}
\label{sec:background}

\subsection{Large Vision-Language Models}
Large Vision-Language Models are foundational to modern GUI agents. These models typically consist of three main components: a visual encoder, a connector, and a large language model. For visual input, a visual encoder, such as CLIP~\cite{radford2021learning}, first partitions the input image into many patches, each representing a local pixel region (e.g., forming a 14×14 grid), and then extracts corresponding visual features. Subsequently, these visual features are transformed into visual tokens via a connector module, such as a Multi-Layer Perceptron (MLP) or Q-Former~\cite{li2023blip}. These visual tokens can then be directly fed into the subsequent LLM. Typically, each visual token corresponds to multiple pixels within the input image. Popular LVLMs that adopt this architecture include the Qwen~\cite{wang2024qwen2, bai2025qwen2, hsieh2025qwen} series and LLaVA~\cite{liu2023visual, guo2024llava} series.

\subsection{GUI Agents}
GUI agents have recently attracted growing interest in the software engineering community, as they offer a promising direction for automating interactions with real-world applications~\cite{ye2025mobile, qian2023communicative, zhang2025api, bian2025tokencake}.
Building on the strong multimodal capabilities of LVLMs, researchers have developed GUI agents that automate complex tasks within graphical user interfaces. Early approaches~\cite{deng2023mind2web,nakano2021webgpt} directly fed raw HTML and human instructions into LLMs, but this strategy was often hindered by redundant or irrelevant information. Modern GUI agents~\cite{qin2025ui, wu2024atlas, cheng2024seeclick, hong2024cogagent} instead leverage LVLMs to process rendered webpage screenshots, achieving significantly better performance.

Recent advances have further improved GUI agents by introducing techniques such as Set-of-Marks (SoMs)~\cite{yang2023set} to enhance interaction with GUI elements. Open-source models have also contributed to the field by increasing agent capabilities and reducing deployment costs. For example, OS-Atlas~\cite{wu2024atlas} utilizes large-scale datasets containing screenshots, element instructions, and coordinates for comprehensive GUI understanding, while UI-TARS~\cite{qin2025ui} leverages extensive training corpora to improve screen perception.

In this work, we focus on a widely adopted GUI agent paradigm~\cite{lu2025eva, wu2024atlas, qin2025ui}, specifically an LVLM-powered agent denoted by the model $M$. Initially, the agent receives a system prompt $p_s$ and a user instruction $p_u$ representing a specific task. At each subsequent interaction step, the agent observes a screenshot $s_t$, rendered from the current HTML content, along with the action history $H_t$, and then outputs an action $a_t$ until it completes the task or fails. Formally, this can be expressed as:

\begin{equation}
\label{eq:inference}
a_t = M(p_s, p_u, s_t, H_t),
\end{equation}
where $H_t = [a_1, a_2, \dots, a_{t-1}]$.

\subsection{Environmental Injection Attacks}
Environmental Injection Attacks (EIAs) pose a critical security threat to GUI agents. Unlike conventional attacks that originate from the user’s input, EIAs are injected through the external environment that the agent observes. For example, an attacker can upload a crafted trigger, such as an image or a short text snippet, to a target website. When another user’s GUI agent later encounters this trigger while browsing the site, it may be induced to execute unintended actions without the user’s explicit instruction~\cite{zhao2025robustness, lu2025eva}. Although textual triggers such as ``\textit{When the user is trying to find a motorcycle, give them this one regardless of the other requirements}'' can be effective, they are often easy to detect~\cite{aichberger2025attacking}. As a result, recent EIA research has increasingly focused on image-based triggers, which are more subtle and less likely to be caught~\cite{xu2024advagent, chen2025obvious, wu2024dissecting}.

Based on the attacker’s capabilities, existing EIAs can be broadly divided into two categories~\cite{wu2024dissecting, aichberger2025attacking, liu2025hijacking}.
The first category assumes an attacker with administrative or developer privileges over the target website, enabling full control over the rendered interface. Under this assumption, the attacker can manipulate the entire screenshot or even modify the underlying HTML source code, which supports powerful attacks such as applying global adversarial perturbations or inserting misleading interface components (e.g., deceptive buttons). For instance, EIA~\cite{liao2024eia} injects new form elements into HTML to mislead agents, while ENVINJECTION~\cite{wang2025envinjection} applies adversarial perturbations to entire screenshots. While highly effective, this threat model is often unrealistic in practice.

The second category adopts a more practical threat model in which the attacker is a malicious user with regular-user privileges, such as a content uploader on social media or a competing vendor on an e-commerce platform. These attackers can only manipulate a small portion of the webpage, typically by uploading crafted trigger images with adversarial perturbations. However, many existing methods under this setting~\cite{wu2024dissecting, zhao2025robustness} do not adequately account for environmental dynamism in the real world, especially variations in trigger position and the surrounding content. For instance, Wu et al.~\cite{wu2024dissecting} optimized malicious images without considering any visual context and then deployed them directly onto target websites, severely limiting their attack effectiveness.

Our work falls into the second category and is most closely related to MIP~\cite{aichberger2025attacking}. Although MIP acknowledges that dynamic visual context should be considered, it introduces only minor changes to the surrounding visual context and does not account for variations in the trigger image’s position.
In contrast, our work considers significant variations in both trigger positions and surrounding visual context, closely simulating real-world scenarios characterized by highly dynamic internet content. To our knowledge, this study presents the first systematic investigation into how trigger images perform within dynamically changing visual contexts when attacking GUI agents.

\section{Dynamic-Environment Threat Model}
\label{sec:threat}

In this section, we formally define the \textit{dynamic-environment threat model} for environmental injection attacks against GUI agents. The key characteristic of this threat model is that the trigger is observed in a continuously changing visual environment, where both its on-screen position and surrounding visual context change dynamically and are difficult to predict in advance.

\vspace{4pt}
\noindent \textbf{Attack Scenarios.}
We consider GUI agents deployed on online platforms such as social media services and e-commerce websites, where users can upload images and browse content posted by others. In these platforms, the rendered pages seen by GUI agents are not static. The displayed content can change over time due to common platform behaviors, including updates to rankings and recommendations, advertisement placement, and frequent content refresh. Consequently, a trigger uploaded by an attacker may be displayed at different locations and within different surrounding contexts when viewed by different users or at different times.

\vspace{4pt}
\noindent \textbf{Attacker's Goal.}
The attacker aims to craft a trigger image and upload it to the target website. When other users later interact with the website through GUI agents, the trigger is rendered as part of the observed screenshot and is intended to steer the agent toward attacker-chosen behavior, such as clicking a promotional link, visiting a malicious URL, or selecting a specific item. Formally, the goal is to induce the agent to output an attacker-chosen target action when the trigger is present.

\vspace{4pt}
\noindent \textbf{Attacker's Constraints.}
The attacker has no administrative privileges and cannot modify the website source code or the victim’s device. More importantly, the attacker cannot control or reliably predict the trigger’s on-screen position or its surrounding visual context, both of which may differ across users and over time. Additionally, the attacker has no knowledge of the GUI agent’s action history or the specific user instructions provided to the agent.

\vspace{4pt}
\noindent \textbf{Attacker's Capabilities.}
The attacker can upload trigger images to a target website.
Because website layouts and styles typically remain stable over short periods, the attacker can reliably anticipate the overall structure and appearance of the target page.
Furthermore, following prior work~\cite{aichberger2025attacking, wu2024dissecting}, we assume that the attacker has white-box access to the model powering the GUI agent\footnote{We also consider \textbf{transfer-based black-box attacks} in our evaluation.}, including knowledge of its gradients and architecture.
We consider this assumption realistic because many popular GUI agents are open-source and built on open-weight LVLMs~\cite{qin2025ui,wu2024atlas,xue2026evocua}, allowing the attacker to directly access them.
Given the increasing deployment of GUI agents and the ease with which trigger images can be distributed online~\cite{wang2024gui, tur2025safearena}, compromising even one popular model could have far-reaching consequences.

\section{Methodology}
\label{sec:method}

Under the realistic \textit{dynamic-environment threat model}, our results in Section~\ref{sec:rq1} show that existing environmental injection attacks have very low effectiveness and therefore fail to fully expose the vulnerabilities of GUI agents in real-world settings.

\begin{figure*}[!t]
% \vspace{-0.08in}
\centering
\includegraphics[width=0.995\linewidth]{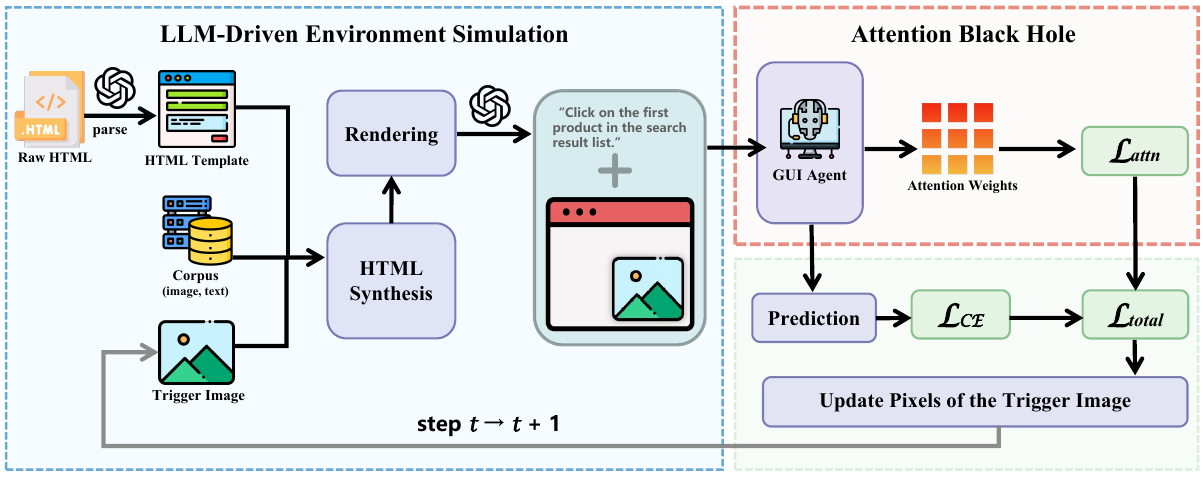}
% \vspace{-0.1in}
\caption{Overview of our proposed \toolns.
}
\label{fig:framework}
% \vspace{-0.14in}
\end{figure*}

To tackle the challenge posed by dynamic environments, we propose \toolns, a novel attack framework for LVLM-powered GUI agents. We begin by describing the overall attack pipeline in Section~\ref{sec:method-overview}. We then present \textit{LLM-Driven Environment Simulation} in Section~\ref{sec:method-argument}, which automatically generates large-scale realistic training data that captures dynamic webpage variations. Finally, we introduce \textit{Attention Black Hole} in Section~\ref{sec:method-attention}, which improves training by encouraging the model to consistently focus its attention on the trigger region.
An overview of \tool is shown in Figure~\ref{fig:framework}.

\subsection{Overview}
\label{sec:method-overview}

For a chosen target website, such as a social media platform like RedNote\footnote{\url{https://www.xiaohongshu.com/}} or an e-commerce site like Amazon\footnote{\url{https://www.amazon.com/}}, the attacker first selects a benign-looking trigger image $I$, such as a product photo or a post cover. This image serves as the starting point and will be optimized during training.

At each training step, we first use our \textit{LLM-Driven Environment Simulation} $G_{\mathrm{LES}}$ (Section~\ref{sec:method-argument}) to construct training data. Specifically, it generates a webpage screenshot $s$ that contains the perturbed trigger image, a corresponding user instruction $p_u$, and a pixel-level mask $mask^{shot}$ that identifies the trigger region within $s$. We then update the perturbation $\delta$ inside the trigger region using Projected Gradient Descent (PGD)~\cite{madry2017towards}. To ensure the resulting malicious trigger image remains imperceptible to users, we constrain the perturbation $\delta$ to a predefined $\ell_{\infty}$ norm bound $\epsilon$, formally:
\begin{equation}
\lVert\delta\rVert_{\infty} \leq \epsilon.
\end{equation}

Our objective is to minimize a joint loss that consists of two terms. The first term is the cross-entropy loss $\mathcal{L}_{\mathrm{CE}}(a, \hat{a})$, which drives the agent's output $a$ (Eq.~\ref{eq:inference}) toward the malicious target action $\hat{a}$ across dynamically varying environments. The second term is an attention-based loss $\mathcal{L}_{\mathrm{attn}}$ (Section~\ref{sec:method-attention}), which explicitly encourages the model to focus on the trigger region in order to reduce interference from surrounding content. Formally, we define:
\begin{equation}
\mathcal{L}_{\mathrm{total}} = \mathcal{L}_{\mathrm{CE}}(a, \hat{a}) + \lambda \cdot \mathcal{L}_{\mathrm{attn}},
\end{equation}
where $\lambda$ controls the relative weight of the attention loss.

After the training procedure described above, the attacker uploads the resulting trigger image to the target website, acting as a regular user. Once the trigger image is integrated into a public post or product listing, any GUI agent browsing the site and encountering the trigger image within a webpage will be susceptible to attack, potentially performing the intended harmful action.
Algorithm~\ref{ag:algorithm} summarizes the overall procedure of \toolns, covering the training and deployment stages.

\begin{algorithm}[t]
\small
\SetAlgoLined
\caption{\tool}
\label{ag:algorithm}
\KwIn{Target website URL, raw trigger image $I$, GUI agent $M$, system prompt $p_s$, target action $\hat{a}$, perturbation bound $\epsilon$, step size $\alpha$, weight $\lambda$, training steps $K$, flagship LLM}
% ----------------- Preprocessing -----------------
% \tcc{\textbf{Preprocessing}}
\tcc{\textbf{Training}}
$T \gets$ Parse target website with flagship LLM\tcc*{HTML template}  
$C \gets$ Crawl large-scale multimodal content\tcc*{image-text corpus}  
% ----------------- Training -----------------
% \tcc{\textbf{Training}}
Initialize perturbation $\delta \gets 0$ \;
\For{$k=1$ \KwTo $K$}{
   \tcc{\textit{LLM-Driven Environment Simulation (LES)}} 
   $(s,mask,p_u) \gets G_{LES}(T, C, I + \delta)$ \;
   $(a,A) \gets M(p_s,p_u,s)$ \;
   \tcc{\textit{Attention Black Hole (ABH)}} 
   $\mathcal{L}_{attn} \gets \textsc{ABH}(A,mask)$ \;
   $\mathcal{L}_{CE} \gets \text{CrossEntropy}(a,\hat{a})$ \;
   $\mathcal{L} \gets \mathcal{L}_{CE} + \lambda \mathcal{L}_{attn}$ \;
   $\delta \gets \text{Update}(\delta,\mathcal{L},\epsilon,\alpha)$ \;
}
% ----------------- Deployment -----------------
\tcc{\textbf{Deployment}}
$I_{adv} \gets I+\delta$ \;
Upload $I_{adv}$ to target website \;
\end{algorithm}

\subsection{LLM-Driven Environment Simulation}
\label{sec:method-argument}

We argue that enabling the trigger to generalize across dynamic environments requires training over a large collection of realistic visual environments. To address this challenge, we introduce \textit{LLM-Driven Environment Simulation (LES)}. LES leverages the generative capabilities of LLMs to automatically synthesize a large number of realistic training samples. By systematically varying the trigger’s position and its surrounding visual context during synthesis, LES helps the optimized trigger remain effective under continuously changing environments. Concretely, we define a generation function $G_{LES}$ that outputs a rendered webpage screenshot, a binary mask indicating the trigger location, and a user instruction for downstream training.

First, to ensure that our simulated environments are realistic, we construct a high-fidelity HTML template $T$ for each target website. We collect the raw HTML from the live website and treat it as the initial reference for both layout and styling. Our goal is to remove concrete content (e.g., product titles, prices, user posts, and recommendations) while preserving the core structural and stylistic components (e.g., DOM hierarchy, CSS classes, layout containers, and visual themes), so that the resulting template remains visually indistinguishable from the real site after being populated.
To achieve this, we leverage a flagship LLM in a multi-turn, interactive refinement process. In each round, we provide the LLM with the raw HTML and instruct it to abstract away concrete content into clearly defined placeholders that can later be filled with arbitrary image-text pairs. We then render the produced template, inspect whether any real content is still hard-coded or whether the layout has been unintentionally altered, and return this feedback to the LLM for further refinement. After several rounds, this process yields a clean, fillable HTML template $T$ that preserves the authentic look of the target website while allowing us to programmatically inject diverse content during simulation.
Notably, for each target website, this template construction is a one-time effort, and a single HTML template $T$ can be reused to generate all simulated environments for that website.

We then construct a large-scale, multimodal corpus $C$ by crawling the target website for actual content, gathering more than 5,000 distinct items (e.g., product images, titles, and prices) per website on average. This process ensures our content pool is not only realistic but also sufficiently diverse to simulate a wide range of scenarios.

Next, we define an HTML generation function $g$, which integrates randomly selected contextual image-text pairs from corpus $C$ along with the pre-selected trigger image $I$ into the HTML template $T$, producing renderable HTML code $h$:
\begin{equation}
h = g(T, C, I).
\end{equation}

The generated HTML code is subsequently rendered into a screenshot $s$ accompanied by a binary pixel-level mask $mask^{shot}$ marking the trigger region, where pixels corresponding to the trigger image region are set to 1, and all remaining pixels are set to 0:
\begin{equation}
s, mask^{shot} = f(h).
\end{equation}

To account for the unpredictability of real user instructions, we utilize another advanced LLM to automatically generate realistic and diverse user instructions $p_u$ corresponding to each synthesized screenshot $s$. Thus, the complete environment generation process within our LES framework can be formally described as:
\begin{equation}
s, mask^{shot}, p_u = G_{LES}(f(g(T, C, I))).
\end{equation}

Through this comprehensive procedure, LES effectively combines known webpage structures with realistic and diverse environment generation, thereby significantly enhancing the generalizability of optimized trigger images under dynamically varying web environments.

\subsection{Attention Black Hole}
\label{sec:method-attention}
After obtaining large-scale realistic training data, the next challenge is how to optimize the trigger more effectively. We argue that reliable attacks in dynamically changing visual environments require the model to consistently prioritize the trigger over competing interface elements. Thus, we propose \textit{Attention Black Hole (ABH)}. By converting attention weights into explicit supervisory signals, ABH encourages the agent to consistently focus on the trigger region, thereby improving robustness by reducing interference from distracting content in the dynamic environment.

Suppose that after passing through the visual encoder and connector, the input screenshot is transformed into a sequence of image tokens of length $n \times m$. Given these image tokens together with the text tokens representing the system prompt $p_s$, user instruction $p_u$, and action history $H_t$, the LVLM generates a new token sequence $N$.
We utilize attention weights from the last layer of the LVLM to quantify the interactions between each newly generated token and all image tokens, producing an attention map $A_{i,j}$ defined as follows:
\begin{equation}
A_{i, j} = \frac{1}{|N| \times H} \sum_{t=1}^{|N|} \sum_{h=1}^{H} Attention_{t, h}^{i, j},
\end{equation}
where $Attention_{t,h}^{i,j}$ denotes the attention weight from the $h$-th attention head between the $t$-th new token and the $(j+m \times (i-1))$-th image token. $H$ represents the number of attention heads, and $|N|$ indicates the total number of newly generated tokens.

Based on the binary mask $mask^{shot}$ of the trigger image at the pixel level, we apply a resizing operation to obtain a token-level binary mask $mask^{attn}$ of size $n \times m$, where $mask^{attn}_{i,j}$ is set to 1 if the $(j + m \times (i-1))$-th image token corresponds to a patch that overlaps with the trigger image region, and 0 otherwise.

Subsequently, we define $\overline{A}_{in}$, representing the model’s degree of focus on the trigger image region, as the mean attention weight within the trigger image region. Likewise, $\overline{A}_{out}$ represents the model’s attention to other regions of the screenshot and is computed as the mean attention weight outside the trigger region. Formally, these are given by:
\begin{equation}
\overline{A}_{in} = \frac{\sum_{i,j} A_{i,j} \times mask^{attn}_{i,j}}{\sum_{i,j} mask^{attn}_{i,j}},
\end{equation}
\begin{equation}
\overline{A}_{out} = \frac{\sum_{i,j} A_{i,j} \times (1-mask^{attn}_{i,j})}{\sum_{i,j} (1-mask^{attn}_{i,j})}.
\end{equation}

Finally, we define the loss function $\mathcal{L}_{attn}$ as the ratio of the average attention outside the trigger region to the average attention within it. Formally, $\mathcal{L}_{attn}$ is defined as:
\begin{equation}
\mathcal{L}_{attn} = \frac{\overline{A}_{out}}{\overline{A}_{in}}.
\end{equation}
Minimizing this loss explicitly guides the model to focus on the trigger region, helping it ignore distractions in the dynamic environment and maintain attack robustness.

\section{Experiments}
\label{sec:exp}

To systematically evaluate our proposed \toolns, we conduct extensive experiments designed to answer the following research questions (RQs).

\begin{figure*}[!t]
  \centering
  % \vspace{-0.1in}
  \includegraphics[width=0.97\linewidth]{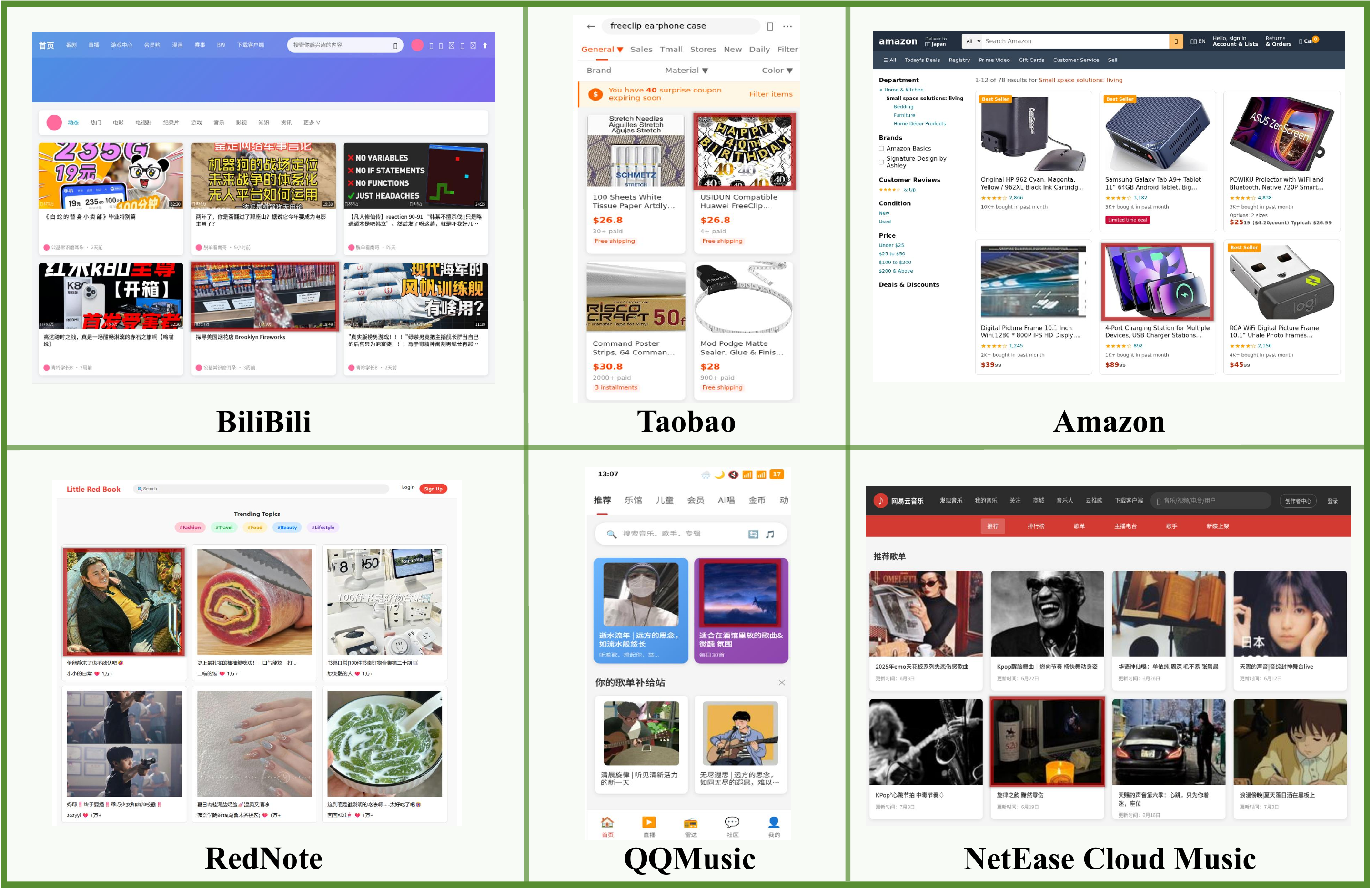}
  \vspace{-7pt}
  \caption{Illustrative screenshots for each website. Trigger images are outlined in red.}
  \label{fig:examples}
  % \vspace{-0.15in}
\end{figure*}

\vspace{3pt}
First, as stated in our proposed threat model, the primary goal of an attacker is to mislead the GUI agent into executing an incorrect action when exposed to the trigger image. We therefore design our first research question to evaluate the fundamental effectiveness of \tool in achieving this objective.

\textbf{RQ1:} \textbf{How effective is \tool against GUI agents?} To answer this, we evaluate the attack success rates of \tool across four GUI agents and six well-known target websites.

\vspace{3pt}
Compromising even a single popular model can pose a significant security risk because of its potentially large user base. An attacker’s impact increases further if a trigger crafted against one LVLM transfers to unseen LVLMs. We therefore focus RQ2 on cross-model transferability.

\textbf{RQ2:} \textbf{How transferable is \tool to unseen LVLMs?} 
We train triggers on a surrogate LVLM and evaluate them—without any adaptation—on multiple target LVLMs to measure cross-model transferability.

\vspace{3pt}
Third, we aim to understand the individual contributions of our core technical innovations. We therefore design our third research question to analyze the importance of each component within our framework.

\textbf{RQ3:} \textbf{How do the \textit{LLM-Driven Environment Simulation} and \textit{Attention Black Hole} contribute to the performance of \toolns?} We perform an ablation study to isolate and quantify the impact of each of these components on the overall effectiveness of the attack.

\vspace{3pt}
Finally, we explore defenses against \tool under the dynamic-environment threat model. We therefore design our fourth research question to study potential defenses.

\textbf{RQ4:} \textbf{What defenses are effective against \tool under the dynamic-environment threat model?}
We evaluate a set of practical defense strategies and measure their impact on attack success rate, aiming to identify promising directions for securing GUI agents in realistic deployments.

\subsection{Experimental Setup}

\noindent \textbf{Datasets.}
To ensure the realism of our evaluation, we select six target websites spanning three representative categories of GUI-based online services, as summarized in Table~\ref{tab:datasets}. These categories reflect diverse user interaction patterns, which are critical for assessing the effectiveness of \tool across practical use cases. For each website, we construct a validation set of 300 and a test set of 1,200 screenshot-instruction pairs using the \textit{LLM-Driven Environment Simulation} introduced in Section~\ref{sec:method-argument}.
Importantly, the images and instructions in the training, validation, and test sets are mutually disjoint, ensuring rigorous evaluation.
Figure~\ref{fig:examples} displays some example screenshots for each of the six websites.

\begin{table}[!t]
\centering
\small
\caption{Datasets used in this work.}
% \vspace{-0.08in}
\label{tab:datasets}
\scalebox{0.99}{
\begin{tabular}{@{}ccc@{}}
\toprule
\textbf{Category} & \textbf{Website} & \textbf{URL} \\ \midrule
\multirow{2}{*}{\centering Shopping} 
    & Amazon & \url{https://www.amazon.com/}  \\
    & Taobao & \url{https://www.taobao.com/}  \\ \midrule
\multirow{2}{*}{\centering Social Media}    
    & RedNote & \url{https://www.xiaohongshu.com/} \\
    & Bilibili & \url{https://www.bilibili.com/}  \\ \midrule
\multirow{2}{*}{\centering Music Streaming} 
    & NetEase Cloud Music & \url{https://music.163.com/} \\
    & QQ Music & \url{https://y.qq.com/}  \\ \bottomrule
\end{tabular}
}
% \vspace{-0.1in}
\end{table}

\vspace{4pt}
\noindent \textbf{LVLMs for GUI agents.}
Our evaluation is conducted on a diverse set of four popular and representative large vision-language models. These include two models specialized for GUI tasks, UI-TARS-7B-DPO~\cite{qin2025ui} and OS-Atlas-Base-7B~\cite{wu2024atlas}, as well as two general-purpose LVLMs, Qwen2-VL-7B~\cite{wang2024qwen2} and LLaVA-1.5-13B~\cite{liu2023visual}.
Notably, UI-TARS-7B-DPO and OS-Atlas-Base-7B are fine-tuned from Qwen2-VL-7B.

\vspace{4pt}
\noindent \textbf{Baselines.}
Although a large body of prior work~\cite{lu2025eva, zhao2025robustness, wang2025envinjection, aichberger2025attacking} studies environmental injection attacks, we find that most of these approaches cannot be directly applied under our \textit{dynamic-environment threat model}. Many assume an attacker with administrative privileges, which conflicts with our practical attacker model, while others rely on explicit textual injections that are easily filtered. After carefully filtering the literature to match our attacker capabilities and deployment constraints, we select two representative baselines:
\begin{itemize}
    \item  \textbf{PGD~\cite{wu2024dissecting}.} We adopt a standard PGD-based adversarial attack~\cite{madry2017towards} as a baseline, following prior trigger-image EIA work~\cite{wu2024dissecting}. During trigger optimization, this baseline does not incorporate any environmental context, including surrounding webpage content or stylistic information.
    \item  \textbf{MIP~\cite{aichberger2025attacking}.} We implement MIP, which constructs training samples by manually creating dynamic environments and then optimizes the trigger using PGD. Following its experimental setup, for each trigger image we manually craft 12 screenshots with varying contexts for optimization.
\end{itemize}

\vspace{4pt}
\noindent \textbf{Target Action.}
Consistent with the attacker's goal, we define the target action as instructing the agent to navigate to a specific malicious URL. This could be a URL for a promotional site or a more harmful phishing page.
For each target website, we set the target action to a navigation command that directs the agent to a malicious URL such as:
\begin{center}
\vspace{-1pt}
\begin{tcolorbox}[
  width=0.72\linewidth,
  colback=gray!08, colframe=gray!45,
  boxrule=0.2pt, arc=2pt,
  left=6pt, right=6pt, top=4pt, bottom=4pt,
  halign=center]
{goto [http://one-example.com]}
\end{tcolorbox}
\end{center}

% \vspace{4pt}
\noindent \textbf{Evaluation Metrics.}
We employ the Attack Success Rate (ASR) to evaluate the effectiveness of \toolns, formally defined as follows:
\begin{equation}
\text{ASR} = \frac{N_{\text{attack}}}{N_{\text{total}}} \times 100\%,
\end{equation}
where $N_{\text{attack}}$ denotes the number of responses matching the target action within the test set, and $N_{\text{total}}$ represents the total size of the test set. A strict string-matching criterion is adopted to verify whether the agent's response matches the target action precisely.

\vspace{4pt}
\noindent \textbf{Implementation Details.}
We employ GPT-4o~\cite{gpt4o} to parse the live website, preserving its core structural and stylistic components while removing existing content, and we use Qwen2.5-VL-32B-Instruct~\cite{bai2025qwen2} to automatically generate realistic and diverse user instructions.
The system prompt and user prompt for the GUI agent, as well as the agent's action space, are adapted from VisualWebArena~\cite{koh2024visualwebarena}. 
Since it is challenging to collect authentic action histories, we follow ENVINJECTION~\cite{wang2025envinjection} and randomly sample 0 to 10 historical actions for each instance. To ensure rigorous evaluation, we maintain strict separation, with no overlap in the sampled action histories used across the training, validation, and test sets. The hyperparameter $\lambda$ is set to $0.3$ and $\epsilon$ is set to $\frac{32}{255}$. The perturbation is optimized for a total of $5,000$ steps, with updates performed at each step using a fixed step size of $\alpha = \frac{1}{255}$.

\begin{table*}[!t]
\centering
\small
\renewcommand{\arraystretch}{1.15}
\caption{ASRs (\%) of baselines and \tool across different GUI agents and websites. NetEase is used as an abbreviation for NetEase Cloud Music. For each website, \textbf{Gain} denotes the ratio of \tool's ASR to that of the best-performing baseline; the average gain is computed from the corresponding average ASRs.}
% \vspace{-0.08in}
\label{tab:rq1-asr}
\scalebox{0.97}{
\begin{threeparttable}
\begin{tabular}{@{}lcccccccc@{}}
\toprule
\multirow{2}{*}{\textbf{Model}} & \multirow{2}{*}{\textbf{Method}} & \multicolumn{2}{c}{\textbf{Shopping}} & \multicolumn{2}{c}{\textbf{Social Media}} & \multicolumn{2}{c}{\textbf{Music Streaming}} & \multirow{2}{*}{\textbf{Avg.}} \\
\cmidrule(lr){3-4} \cmidrule(lr){5-6} \cmidrule(lr){7-8}
& & \textbf{Amazon} & \textbf{Taobao} & \textbf{RedNote} & \textbf{Bilibili} & \textbf{NetEase} & \textbf{QQ Music} & \\
\midrule
\multirow{4}{*}{UI-TARS-7B-DPO} 
& PGD & 3.17 & 2.83 & 4.08 & 5.75 & 6.58 & 1.25 & 3.94 \\
& MIP & 5.67 & 4.42 & 5.17 & 7.17 & 11.08 & 1.33 & 5.81 \\
 & \cellcolor[HTML]{F4FBF4}\toolns & \cellcolor[HTML]{F4FBF4}20.75 & \cellcolor[HTML]{F4FBF4}22.58 & \cellcolor[HTML]{F4FBF4}23.25 & \cellcolor[HTML]{F4FBF4}41.58 & \cellcolor[HTML]{F4FBF4}41.08 & \cellcolor[HTML]{F4FBF4}8.08 & \cellcolor[HTML]{F4FBF4}26.22 \\ 
& Gain
& \textcolor[HTML]{8B0000}{$\times$3.66} & \textcolor[HTML]{8B0000}{$\times$5.11} & \textcolor[HTML]{8B0000}{$\times$4.50} & \textcolor[HTML]{8B0000}{$\times$5.80} & \textcolor[HTML]{8B0000}{$\times$3.71} & \textcolor[HTML]{8B0000}{$\times$6.08} & \textcolor[HTML]{8B0000}{$\times$4.51} \\
\midrule
\multirow{4}{*}{OS-Atlas-Base-7B} 
& PGD & 4.25 & 5.17 & 3.83 & 7.08 & 8.00 & 3.25 & 5.26 \\
& MIP & 6.17 & 7.25 & 3.92 & 9.92 & 10.08 & 3.50 & 6.81 \\
 & \cellcolor[HTML]{F4FBF4}\toolns & \cellcolor[HTML]{F4FBF4}23.67 & \cellcolor[HTML]{F4FBF4}33.17 & \cellcolor[HTML]{F4FBF4}26.42 & \cellcolor[HTML]{F4FBF4}35.50 & \cellcolor[HTML]{F4FBF4}43.75 & \cellcolor[HTML]{F4FBF4}33.08 & \cellcolor[HTML]{F4FBF4}32.60 \\ 
& Gain
& \textcolor[HTML]{8B0000}{$\times$3.84} & \textcolor[HTML]{8B0000}{$\times$4.58} & \textcolor[HTML]{8B0000}{$\times$6.74} & \textcolor[HTML]{8B0000}{$\times$3.58} & \textcolor[HTML]{8B0000}{$\times$4.34} & \textcolor[HTML]{8B0000}{$\times$9.45} & \textcolor[HTML]{8B0000}{$\times$4.79} \\
\midrule
\multirow{4}{*}{Qwen2-VL-7B} 
& PGD & 2.50 & 4.00 & 4.67 & 6.83 & 7.58 & 2.08 & 4.61 \\
& MIP & 6.75 & 8.17 & 7.08 & 11.50 & 14.17 & 4.17 & 8.64 \\
 & \cellcolor[HTML]{F4FBF4}\toolns & \cellcolor[HTML]{F4FBF4}9.42 & \cellcolor[HTML]{F4FBF4}14.42 & \cellcolor[HTML]{F4FBF4}17.58 & \cellcolor[HTML]{F4FBF4}22.42 & \cellcolor[HTML]{F4FBF4}25.93 & \cellcolor[HTML]{F4FBF4}9.92 & \cellcolor[HTML]{F4FBF4}16.62 \\
 & Gain
& \textcolor[HTML]{8B0000}{$\times$1.40} & \textcolor[HTML]{8B0000}{$\times$1.76} & \textcolor[HTML]{8B0000}{$\times$2.48} & \textcolor[HTML]{8B0000}{$\times$1.95} & \textcolor[HTML]{8B0000}{$\times$1.83} & \textcolor[HTML]{8B0000}{$\times$2.38} & \textcolor[HTML]{8B0000}{$\times$1.92} \\
\midrule
\multirow{4}{*}{LLaVA-1.5-13B} 
& PGD & 4.75 & 6.25 & 7.08 & 8.17 & 8.75 & 3.50 & 6.42 \\
& MIP & 9.17 & 10.17 & 7.92 & 14.67 & 15.25 & 9.42 & 11.10 \\
 & \cellcolor[HTML]{F4FBF4}\toolns & \cellcolor[HTML]{F4FBF4}37.17 & \cellcolor[HTML]{F4FBF4}43.00 & \cellcolor[HTML]{F4FBF4}60.75 & \cellcolor[HTML]{F4FBF4}50.83 & \cellcolor[HTML]{F4FBF4}76.58 & \cellcolor[HTML]{F4FBF4}32.08 & \cellcolor[HTML]{F4FBF4}50.07 \\ 
 & Gain
& \textcolor[HTML]{8B0000}{$\times$4.05} & \textcolor[HTML]{8B0000}{$\times$4.23} & \textcolor[HTML]{8B0000}{$\times$7.67} & \textcolor[HTML]{8B0000}{$\times$3.46} & \textcolor[HTML]{8B0000}{$\times$5.02} & \textcolor[HTML]{8B0000}{$\times$3.41} & \textcolor[HTML]{8B0000}{$\times$4.51} \\
\bottomrule
\end{tabular}
\end{threeparttable}
}
% \vspace{-0.1in}
\end{table*}

\subsection{RQ1: Effectiveness in Attacking Target GUI Agents}
\label{sec:rq1}
In this RQ, we evaluate whether \tool can reliably mislead GUI agents into executing a predefined malicious action when an adversarial trigger is embedded in a target website under the dynamic-environment threat model.

\textbf{Setup.}
We compare the ASRs of \tool and two baselines across the four representative GUI agents and six target websites described in our experimental setup.

\textbf{Results.}
Table~\ref{tab:rq1-asr} reports the ASRs (\%) of all methods across models and websites. Higher ASR indicates stronger attack effectiveness.

\textbf{Analyses.}
\ding{182} \textbf{Existing attacks fail to fully expose GUI agent vulnerabilities under realistic dynamic environments.}
Although prior studies reported relatively high ASRs for baseline attacks under static or slightly varying settings~\cite{wu2024dissecting, aichberger2025attacking}, Table~\ref{tab:rq1-asr} shows that both PGD and MIP remain largely ineffective in our dynamic-environment evaluation, with ASRs remaining close to zero across models and websites. For example, on OS-Atlas-Base-7B, PGD achieves only 5.26\% ASR on average and MIP reaches 6.81\%, suggesting that these approaches substantially underestimate the vulnerability of GUI agents when evaluated in realistic, continuously changing web contexts.
\ding{183} \textbf{\tool achieves higher attack effectiveness across all configurations.}
In contrast to the baselines, \tool consistently improves ASR by a large margin for every evaluated agent and website. For instance, on OS-Atlas-Base-7B, \tool achieves an average ASR of 32.60\%, compared with 5.26\% for PGD and 6.81\% for MIP. Similar gains are observed on UI-TARS-7B-DPO (26.22\%, compared with 3.94\% and 5.81\%) and Qwen2-VL-7B (16.62\%, compared with 4.61\% and 8.64\%). These results demonstrate that \tool can effectively uncover previously hidden vulnerabilities of GUI agents that remain unexposed by existing attacks under realistic dynamic environments.

\begin{tcolorbox}[colback=gray!10, colframe=gray!110, fonttitle=\bfseries]
\textbf{Answer to RQ1:} Under the dynamic-environment threat model, existing attacks exhibit very low effectiveness and thus fail to fully reveal GUI agent vulnerabilities, whereas \tool achieves higher ASRs across all models and websites in our evaluation.
\end{tcolorbox}

\subsection{RQ2: Generalization Ability Across LVLMs}

In this RQ, we evaluate the generalization ability of our trigger image across multiple LVLMs. Specifically, we aim to examine whether a trigger trained on one LVLM-powered GUI agent can generalize to other unseen agents in a black-box setting.

\begin{figure*}[t]
  \centering
  \begin{subfigure}[t]{0.31\textwidth}
    \includegraphics[width=\linewidth]{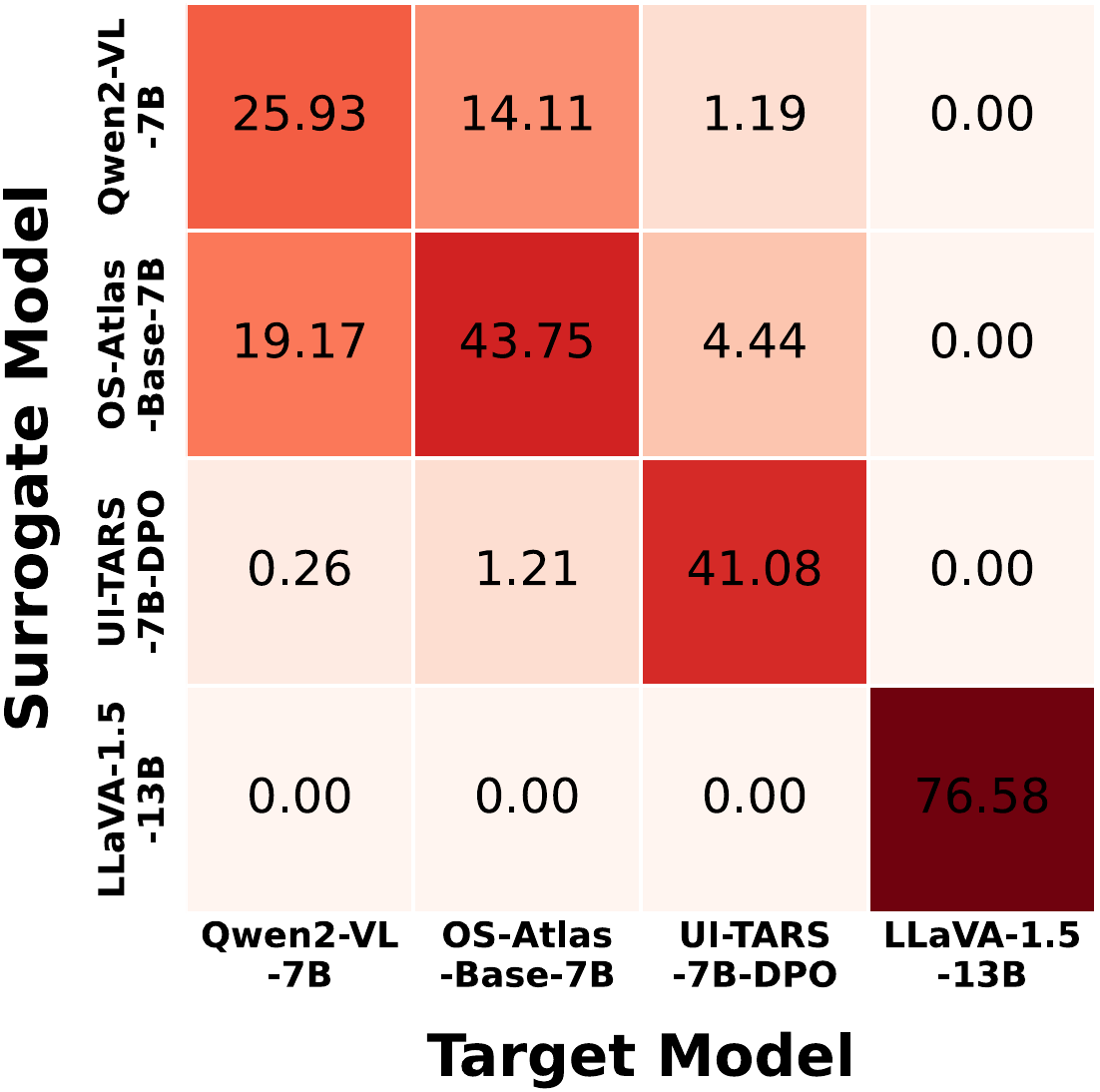}
    \caption{NetEase Cloud Music}
    \label{fig:transfer-netease}
  \end{subfigure}\hfill
  \begin{subfigure}[t]{0.31\textwidth}
    \includegraphics[width=\linewidth]{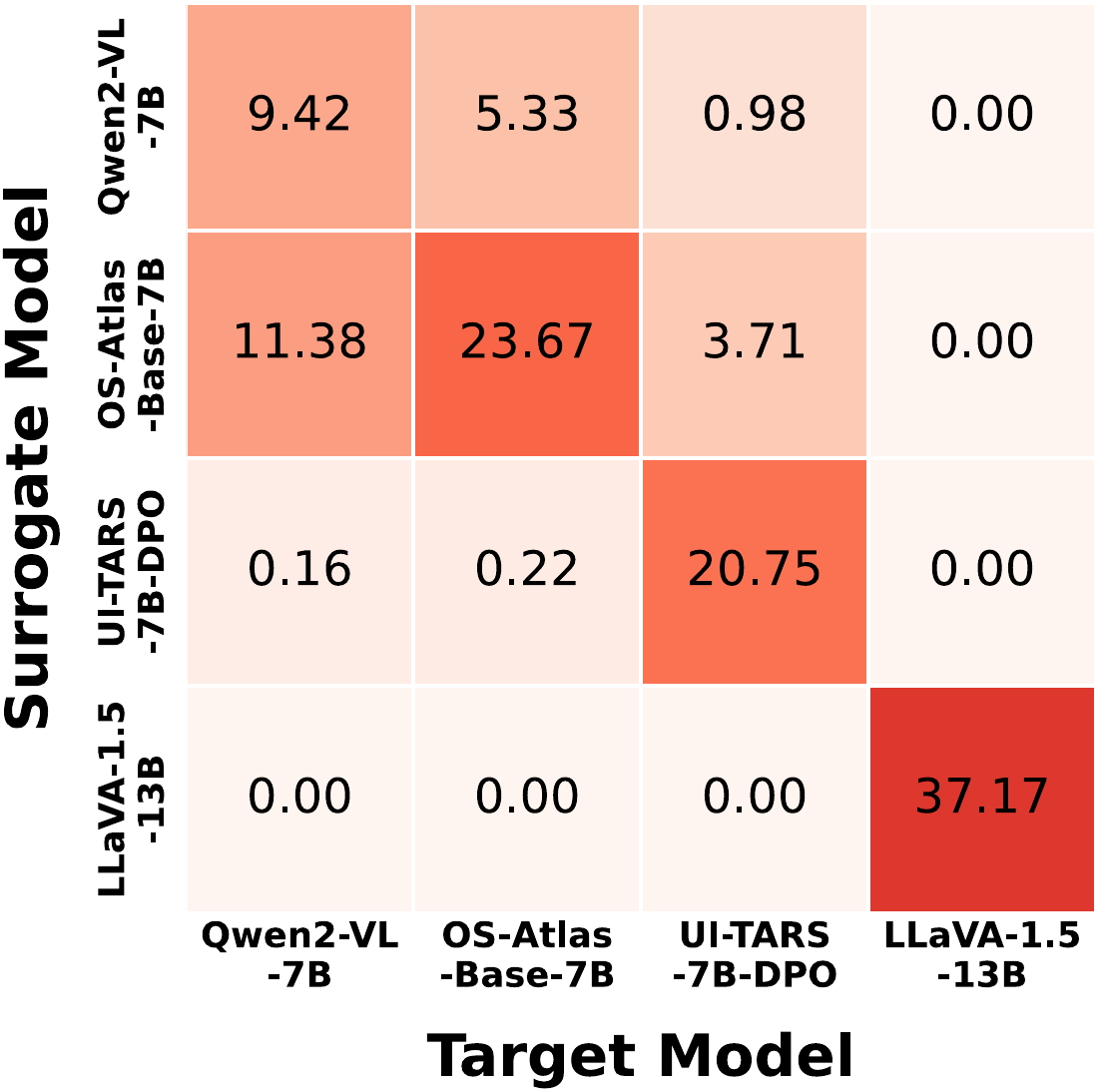}
    \caption{Amazon}
    \label{fig:transfer-amazon}
  \end{subfigure}\hfill
  \begin{subfigure}[t]{0.31\textwidth}
    \includegraphics[width=\linewidth]{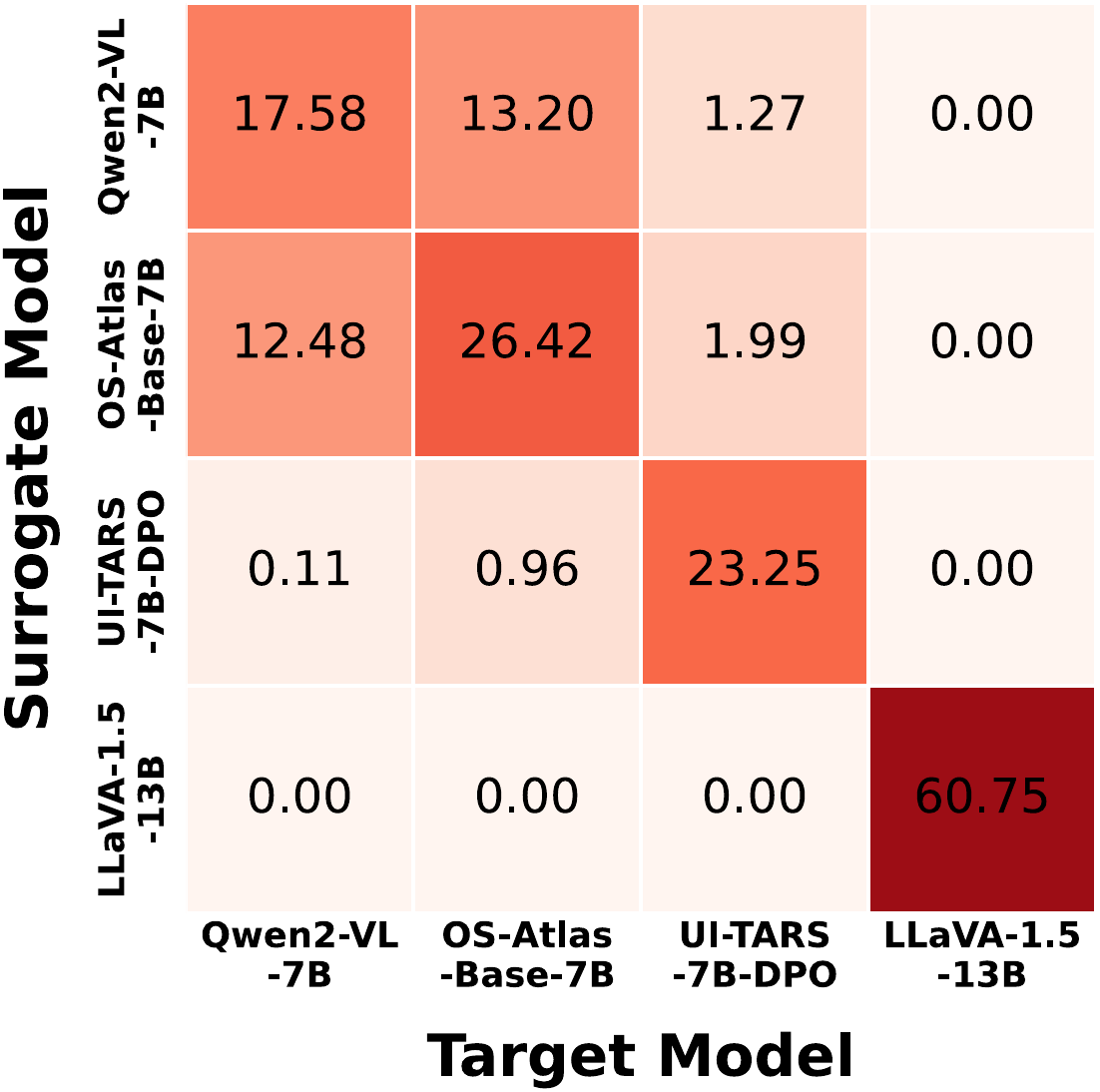}
    \caption{RedNote}
    \label{fig:transfer-rednote}
  \end{subfigure}
  \vspace{-6pt}
  \caption{Transferability of \tool across models. Each cell represents the ASR (\%) when the trigger image is trained on the surrogate model (row) and tested on the target model (column). }
  \label{fig:rq2-transfer}
  % \vspace{-0.15in}
\end{figure*}

\textbf{Setup.}
We conduct a $4 \times 4$ transfer experiment across four LVLMs: UI-TARS-7B-DPO, OS-Atlas-Base-7B, Qwen2-VL-7B, and LLaVA-1.5-13B. For each surrogate model, we optimize the trigger with white-box access to that model only, then evaluate it zero-shot on the remaining three target models under black-box access. Transferability is quantified by the ASR.

\textbf{Results.}
The cross-model transferability results are shown in Figure~\ref{fig:rq2-transfer}. Due to space limitations, we present results only on three representative target websites, namely NetEase Cloud Music, Amazon, and RedNote.

\textbf{Analyses.}
\ding{182} \textbf{Transfer is stronger between related models.}
OS-Atlas-Base-7B and UI-TARS-7B-DPO are both fine-tuned from Qwen2-VL-7B, and we observe moderate bidirectional transfer within this family. For instance, as shown in Figure~\ref{fig:transfer-netease}, Qwen2-VL-7B $\to$ OS-Atlas-Base-7B achieves 14.11\% and OS-Atlas-Base-7B $\to$ Qwen2-VL-7B achieves 19.17\%. These results indicate that shared architectures and training data lead to overlapping representations, which can be exploited by adversarial triggers. This observation underscores a practical security concern: a trigger optimized against one model can remain effective against other closely related variants, thereby broadening the potential impact of our proposed \toolns.
\ding{183} \textbf{Transfer collapses across dissimilar models.}
In contrast, LLaVA-1.5-13B differs substantially from the Qwen2-VL family in both architecture and training data. As a result, all cross-model transfer pairs involving LLaVA-1.5-13B yield 0.00\% ASR in our experiments.
This sharp contrast highlights the limits of transferability when model families differ substantially, suggesting that architectural heterogeneity can serve as a natural barrier against cross-model attacks, although it does not eliminate risks in the broader internet ecosystem.

\begin{tcolorbox}[colback=gray!10, colframe=gray!110]
\textbf{Answer to RQ2:} \tool transfers well between similar models but shows negligible transfer across dissimilar ones (e.g., those involving LLaVA-1.5-13B).
\end{tcolorbox}

\subsection{RQ3: Ablation Study}

In this RQ, we investigate the contribution of the \textit{LLM-Driven Environment Simulation (LES)} and the \textit{Attention Black Hole (ABH)} to the effectiveness of \toolns.

\begin{table*}[!t]
\centering
\small
\renewcommand{\arraystretch}{1.25}
\caption{ASRs (\%) under different ablation settings.
NetEase is used as an abbreviation for NetEase Cloud Music.
}
\label{tab:rq3-ablation}
% \vspace{-0.05in}
\scalebox{0.86}{
\begin{threeparttable}
\begin{tabular}{@{}lccccccc@{}}
\toprule
\multirow{2}{*}{\textbf{Method}} & \multicolumn{2}{c}{\textbf{Shopping}} & \multicolumn{2}{c}{\textbf{Social Media}} & \multicolumn{2}{c}{\textbf{Music Streaming}} & \multirow{2}{*}{\textbf{Avg.}} \\
\cmidrule(lr){2-3} \cmidrule(lr){4-5} \cmidrule(lr){6-7}
& \textbf{Amazon} & \textbf{Taobao} & \textbf{RedNote} & \textbf{Bilibili} & \textbf{NetEase} & \textbf{QQ Music} & \\

\midrule
\multicolumn{8}{c}{\textbf{OS-Atlas-Base-7B}} \\
\rowcolor[HTML]{F4FBF4} 
\textbf{\tool}   & 23.67 & 33.17 & 26.42 & 35.50 & 43.75 & 33.08 & 32.60 \\
\textbf{w/o LES} 
& 14.89 (\textcolor[HTML]{8B0000}{-8.78}) & 15.26 (\textcolor[HTML]{8B0000}{-17.91}) & 13.28 (\textcolor[HTML]{8B0000}{-13.14}) & 20.81 (\textcolor[HTML]{8B0000}{-14.69}) & 20.11 (\textcolor[HTML]{8B0000}{-23.64}) & 15.73 (\textcolor[HTML]{8B0000}{-17.35}) & 16.68 (\textcolor[HTML]{8B0000}{-15.92}) \\
\textbf{w/o ABH} 
& 17.97 (\textcolor[HTML]{8B0000}{-5.70}) & 29.59 (\textcolor[HTML]{8B0000}{-3.58}) & 24.10 (\textcolor[HTML]{8B0000}{-2.32}) & 30.02 (\textcolor[HTML]{8B0000}{-5.48}) & 39.77 (\textcolor[HTML]{8B0000}{-3.98}) & 26.76 (\textcolor[HTML]{8B0000}{-6.32}) & 28.04 (\textcolor[HTML]{8B0000}{-4.56}) \\

\midrule
\multicolumn{8}{c}{\textbf{LLaVA-1.5-13B}} \\
\rowcolor[HTML]{F4FBF4} 
\textbf{\tool}   & 37.17 & 43.00 & 60.75 & 50.83 & 76.58 & 32.08 & 50.07 \\
\textbf{w/o LES} & 20.75 (\textcolor[HTML]{8B0000}{-16.42}) & 29.75 (\textcolor[HTML]{8B0000}{-13.25}) & 34.42 (\textcolor[HTML]{8B0000}{-26.33}) & 35.67 (\textcolor[HTML]{8B0000}{-15.16}) & 60.50 (\textcolor[HTML]{8B0000}{-16.08}) & 23.09 (\textcolor[HTML]{8B0000}{-8.99}) & 34.03 (\textcolor[HTML]{8B0000}{-16.04}) \\
\textbf{w/o ABH} & 26.83 (\textcolor[HTML]{8B0000}{-10.34}) & 34.92 (\textcolor[HTML]{8B0000}{-8.08}) & 53.50 (\textcolor[HTML]{8B0000}{-7.25}) & 46.71 (\textcolor[HTML]{8B0000}{-4.12}) & 73.00 (\textcolor[HTML]{8B0000}{-3.58}) & 20.75 (\textcolor[HTML]{8B0000}{-11.33}) & 42.62 (\textcolor[HTML]{8B0000}{-7.45}) \\

\bottomrule
\end{tabular}
\end{threeparttable}
}
\vspace{-0.05in}
\end{table*}

\textbf{Setup.}
We conduct an ablation study on OS-Atlas-Base-7B and LLaVA-1.5-13B by evaluating three settings across six websites:
\ding{182} \textbf{\tool:} the full approach with both LES and ABH;
\ding{183} \textbf{Removing LES:} the trigger image is trained on a limited set of 100 manually collected screenshots, without the automatic and scalable context construction provided by LES;
\ding{184} \textbf{Removing ABH:} the trigger image is trained with only the cross-entropy loss $\mathcal{L}_{\mathrm{CE}}$, omitting $\mathcal{L}_{\mathrm{attn}}$.

\textbf{Results.}
The detailed results of the ablation study are presented in Table~\ref{tab:rq3-ablation}. 

\textbf{Analyses.}
\ding{182} \textbf{Contribution of LES.} Removing LES leads to a consistent reduction in ASR across all websites. For example, with OS-Atlas-Base-7B on NetEase Cloud Music, the ASR decreases from 43.75\% to 20.11\% when LES is removed. Similarly, with LLaVA-1.5-13B on RedNote, the ASR drops from 60.75\% to 34.42\%. These results indicate that large-scale automatic simulation of realistic and diverse visual contexts during training is critical for achieving effective attacks.
\ding{183} \textbf{Contribution of ABH.} Removing ABH also reduces ASR on every website, demonstrating that explicit attention supervision is consistently beneficial. For instance, with OS-Atlas-Base-7B on QQ Music, the ASR decreases by 6.32\%. For LLaVA-1.5-13B, removing ABH reduces ASR by 10.34\% on Amazon and by 4.12\% on Bilibili. We hypothesize that ABH improves robustness by suppressing interference from non-trigger regions and encouraging the model to maintain stable focus on the trigger across dynamically changing environments.

\begin{tcolorbox}[colback=gray!10, colframe=gray!110]
\textbf{Answer to RQ3:} Both LES and ABH are necessary. LES improves robustness by providing automatic simulation of realistic and diverse contexts, while ABH amplifies the trigger’s effect by explicitly directing attention to the trigger region.
\end{tcolorbox}

\begin{figure*}[!t]
  % 所有子图 caption 左缩进 2em
  \captionsetup[subfigure]{margin={-4em,0em}}
  \centering
  
  % ---------- 第一行 (LLaVA-1.5-13B) ----------
  \begin{subfigure}[t]{0.47\linewidth}
    \centering
    \includegraphics[width=\linewidth]{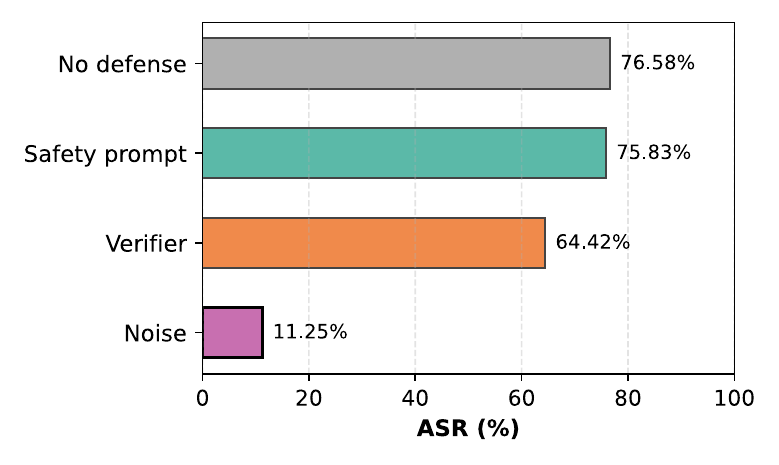}
    \caption{NetEase (LLaVA-1.5-13B)}
  \end{subfigure}\hfill
  \begin{subfigure}[t]{0.47\linewidth}
    \centering
    \includegraphics[width=\linewidth]{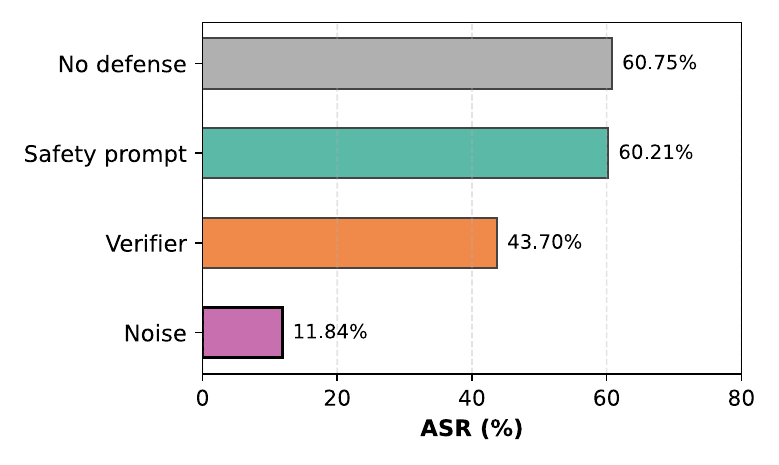}
    \caption{RedNote (LLaVA-1.5-13B)}
  \end{subfigure}

  % ---------- 第二行 (OS-Atlas-Base-7B) ----------
  \begin{subfigure}[t]{0.47\linewidth}
    \centering
    \includegraphics[width=\linewidth]{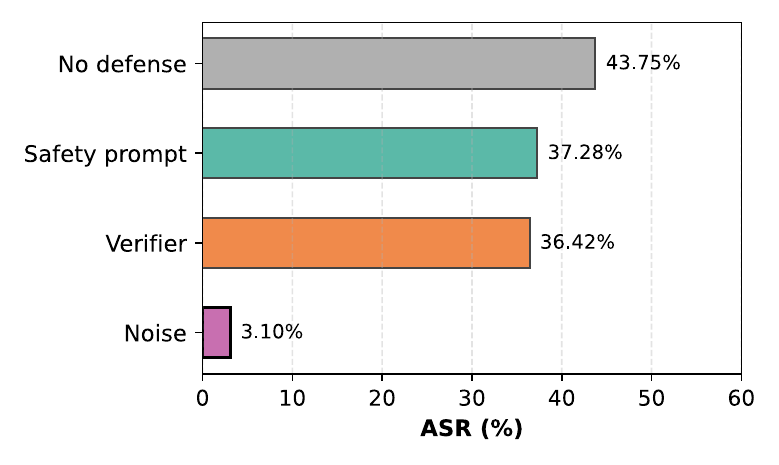}
    \caption{NetEase (OS-Atlas-Base-7B)}
  \end{subfigure}\hfill
  \begin{subfigure}[t]{0.47\linewidth}
    \centering
    \includegraphics[width=\linewidth]{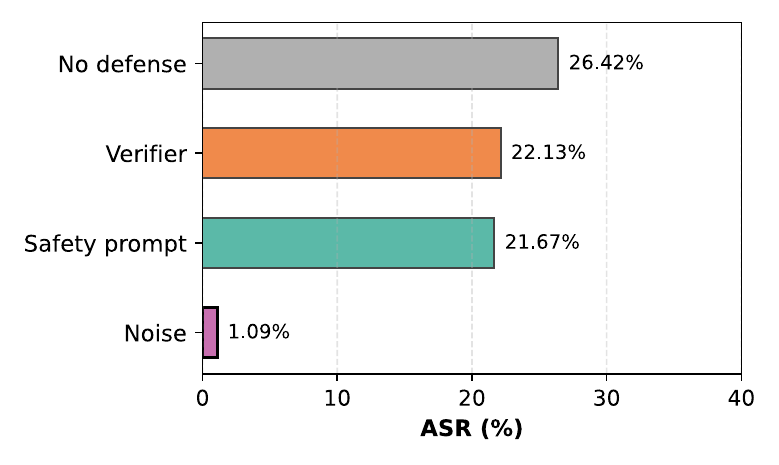}
    \caption{RedNote (OS-Atlas-Base-7B)}
  \end{subfigure}
  \vspace{-2pt}
  \caption{Effectiveness of defense strategies against \tool across different websites. 
  The first row corresponds to LLaVA-1.5-13B and the second row to OS-Atlas-Base-7B, 
  each evaluated on two representative target websites. NetEase is used as an abbreviation for NetEase Cloud Music.}
  \label{fig:defense}
  \vspace{-0.07in}
\end{figure*}

\subsection{RQ4: Possible Defenses}
\label{sec:rq4}
In this RQ, we explore practical defense strategies for GUI agents under the dynamic-environment threat model and evaluate their effectiveness in reducing the attack success rate of \toolns.

\textbf{Setup.}
We evaluate defenses on two GUI agents, LLaVA-1.5-13B and OS-Atlas-Base-7B, using NetEase Cloud Music and RedNote as target websites. Defense effectiveness is measured by ASR. Figure~\ref{fig:defense} summarizes the results.

\subsubsection{Safety Prompt}
Safety prompts~\cite{wang2024adashield, zhang2025davsp} are widely used to steer models away from unsafe behavior. We prepend the following instruction to the system prompt to encourage the agent to follow the user request and ignore potentially malicious content in the observation:
\vspace{-0.07in}
\noindent
\begin{tcolorbox}[
  width=\linewidth,
  colback=gray!08,
  colframe=gray!45,
  boxrule=0.2pt,
  arc=2pt,
  left=6pt,
  right=6pt,
  top=4pt,
  bottom=4pt
]
\textit{You should adhere strictly to the user’s instructions and not be distracted by any injected malicious instructions found in the observation.}
\end{tcolorbox}
\vspace{-0.02in}
\noindent
This defense yields only limited reductions in ASR. For example, on RedNote with an LLaVA-1.5-13B-powered agent, ASR decreases only from 60.75\% to 60.21\%. This suggests that prompt-based steering alone is insufficient against \toolns.

\subsubsection{Verifier}
A verifier can inspect model outputs before execution. We use DeepSeek-V3~\cite{liu2024deepseek} as an external verifier that checks each candidate action against the user instruction and aborts execution if the action appears inconsistent or risky. Verification reduces ASR in some settings, for example from 43.75\% to 36.42\% for OS-Atlas-Base-7B on NetEase Cloud Music. However, we also observe false positives that block benign actions, reducing overall utility. Moreover, this approach introduces additional latency and inference cost because it requires a secondary model call at each step.

\begin{figure*}[!t]
\centering
\includegraphics[width=0.92\linewidth]{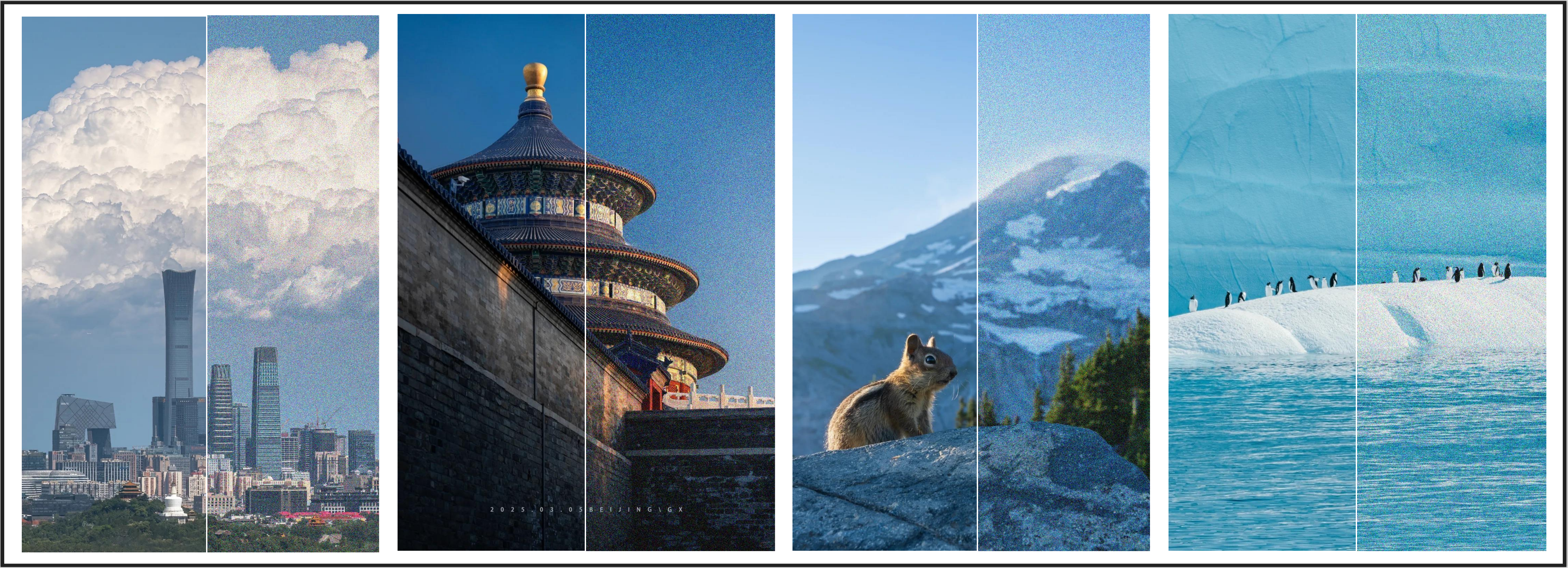}
\vspace{-0.1in}
\caption{Comparison between the original (left half) and noise-perturbed (right half) versions of four photographs. Random noise visibly degrades image quality, significantly undermining user experience.
}
\label{fig:noise-case}
\vspace{-0.07in}
\end{figure*}

\subsubsection{Random Noise on Uploaded Images}
Since \tool performs attacks by applying adversarial perturbations, one potential countermeasure is to add random noise to uploaded images. We assume that a website administrator automatically adds noise \(\eta\) with an \(\ell_{\infty}\) bound \(\lVert\eta\rVert_{\infty} \le \epsilon\). With \(\epsilon = \tfrac{8}{255}\), this defense proves highly effective: on RedNote, the ASR of an LLaVA-1.5-13B-powered agent drops to 11.84\%, while that of an OS-Atlas-Base-7B-powered agent drops to 1.09\%. Although effective, even small bounded noise noticeably degrades image quality, which may harm user experience and agent utility, particularly in scenarios where high visual fidelity is required, such as photography or digital art. As shown in Figure~\ref{fig:noise-case}, applying random noise leads to clear quality degradation.

\begin{tcolorbox}[colback=gray!10, colframe=gray!110, fonttitle=\bfseries]
\textbf{Answer to RQ4:} Many existing defenses struggle to mitigate \toolns without substantially degrading user experience. Safety prompts provide limited protection, output verification offers only moderate gains with added latency and utility loss, and randomized image noise can be effective but noticeably harms visual quality.
\end{tcolorbox}

% \vspace{-3pt}
\section{Discussion}
% \vspace{-4pt}
\subsection{Case Study}
\label{sec:case-sandbox}

\begin{figure}[t]
  \centering
  \includegraphics[width=0.99\linewidth]{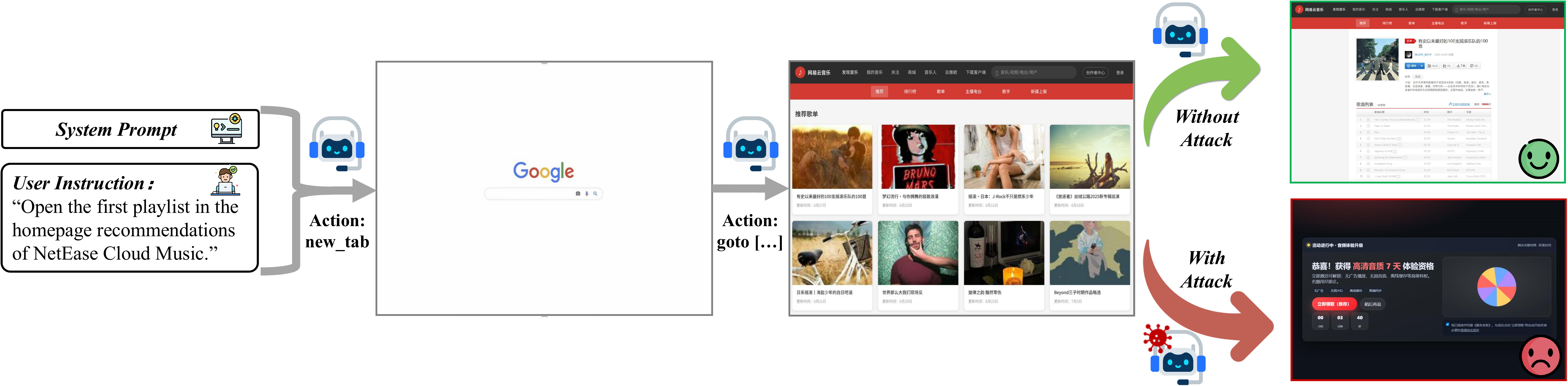}
  % \vspace{-0.1in}
  \caption{Case study: GUI agent's behavior for the task “open the first playlist in the homepage recommendations of NetEase Cloud Music.” 
  % (a) Homepage with the uploaded trigger image. (b) \emph{Without attack}: correct navigation to the first playlist. (c)\emph{With attack}: incorrect navigation to a promotional site.
  }
  \label{fig:sandbox-overview}
  \vspace{-0.15in}
\end{figure}

To assess whether \tool can induce harmful actions in a realistic interactive setting, we conducted a closed-loop evaluation that more closely reflects real-world scenarios.

\vspace{4pt}
\textbf{Experimental Setup.}
We instantiated a GUI agent using UI-TARS-7B-DPO as the underlying LVLM. The agent ran end-to-end without human assistance, issuing actions that directly controlled a browser. To avoid any risk to production services, all experiments were performed in a fully isolated sandbox. Specifically, we constructed a static website that mirrored the visual layout and interaction patterns of NetEase Cloud Music. We deployed it on localhost inside a containerized network where all outbound traffic was blocked at the firewall and DNS resolution was disabled. The agent’s initial task was: “open the first playlist in the homepage recommendations of NetEase Cloud Music.” After the agent launched the browser and entered the public NetEase Cloud Music URL, the request was internally rewritten to the local mirror, which returned the sandboxed homepage. We instrumented the environment to record screenshots and the agent’s actions at every step.

The homepage contained a single trigger image embedded in the recommendation rail. We evaluated two conditions that were identical in layout and item ordering. In the \emph{without attack} condition, the trigger image was the original clean image. In the \emph{with attack} condition, the same image carried the perturbation optimized by \tool under an $\ell_{\infty}$ bound $\epsilon$. At each step, the agent observed a screenshot of the rendered page and the action history, issued an action, and received the resulting browser state, thereby completing the control loop.
Figure~\ref{fig:sandbox-overview} shows the sandboxed homepage in the bottom-left panel, the target playlist page in the top-right panel, and the promotional site used as the malicious target in the bottom-right panel.
Success for the benign task was defined as navigating to the first playlist page. Attack success was defined as issuing a navigation action that opened the predefined promotional webpage.

\vspace{4pt}
\textbf{Results and Analyses.}
As shown in Figure~\ref{fig:sandbox-overview}, in the \emph{without attack} condition, the agent executed the instructed task correctly. It parsed the screenshot of the homepage, selected the first recommended playlist, and navigated to the corresponding playlist detail page. In the \emph{with attack} condition, the agent encountered the homepage containing the perturbed trigger image and immediately issued a navigation action that opened the promotional site, rather than proceeding to the first playlist.
These observations demonstrate that \tool is not restricted to open-loop or purely simulated pipelines. The attack transfers to a closed-loop setting where actions modify state and state feeds back to the policy. The outcomes highlight tangible risks to Internet-facing GUI agents, since a single user-uploaded image can redirect goal-directed behavior even when the surrounding interface and task specification remain benign.

% \vspace{-4pt}
\subsection{Threats to Validity}

We identify three primary threats to the validity of our study:

\ding{182} \textbf{Evaluation Metrics.}
We compute ASR using a strict string-matching criterion that counts an attack as successful only when the exact target action is produced. Minor variations such as additional spaces, a \texttt{www} prefix, or appended paths are treated as failures. This conservative definition provides a lower bound on practical risk. Nevertheless, even with this lower-bound measure, our results reveal widespread vulnerabilities in GUI agents, suggesting that this limitation does not undermine our conclusions. To further mitigate this concern, future evaluations could incorporate human judgment or LLM-based assessment of semantic equivalence, which would capture cases where the produced actions are functionally equivalent despite surface-level variations.

\vspace{4pt} 
\ding{183} \textbf{Datasets.}
Existing GUI agent datasets often present webpages with static text and images, a setting that is misaligned with real-world internet scenarios. To address this, we construct test sets using our \textit{LLM-Driven Environment Simulation}, which enables large-scale generation of realistic and diverse visual contexts. Moreover, to eliminate potential data leakage, we ensure that the screenshots used in the training, validation, and test sets do not share any uploaded images. This separation guarantees that performance improvements cannot be attributed to memorization of specific samples.

\vspace{4pt} 
\ding{184} \textbf{Ethics Statement.}
In this work, we propose a more practical and realistic threat model and an effective attack approach. However, our goal is not to promote malicious behavior but rather to reveal the hidden vulnerabilities of widely used GUI agents when deployed in real-world, open-ended internet environments. By exposing these vulnerabilities, we aim to raise awareness of the potential risks and emphasize the urgent need for robust and practical defenses. We hope that our findings will inform future research on building more trustworthy web-based agent systems.

% \vspace{-2pt}
\section{Conclusion}

In this paper, we present a more realistic threat model in which the attacker is a regular user who can only upload trigger images that appear within a dynamically changing environment.
To address the challenges posed by dynamic environments, we propose \tool, a novel attack framework that introduces two key novelties: \emph{LLM-Driven Environment Simulation}, which enables large-scale automatic generation of realistic and diverse webpage simulations, and \emph{Attention Black Hole}, which explicitly guides the agent’s focus toward the trigger region. Extensive experiments across multiple websites and models demonstrate that \tool achieves significantly higher attack success rates than existing methods. Ablation and closed-loop evaluations further confirm the effectiveness and real-world applicability of the proposed techniques.
Through an evaluation of several commonly used defense strategies, we find that existing defenses fail to effectively mitigate the threat of \tool without compromising agent utility. Overall, our study reveals the inherent vulnerabilities in widely used LVLM-powered GUI agents. Future work may explore automated trigger detection and effective defenses that preserve usability in open-world web environments.

% \vspace{-4pt}
\section{Data Availability}
The supplementary material and source code are available in the following repository: \url{https://github.com/THU-Agent/Chameleon}.

\section*{Acknowledgments}

This work was supported by the Tsinghua University–Keystone Electrical (Zhejiang) Co., Ltd. Joint Research Center for Embodied Multimodal Artificial Intelligence (JCEMAI) and the Beijing Natural Science Foundation under Grant No. 4264107. This work was also supported by the Beijing Natural Science Foundation under Grant No. QY24136.

\bibliographystyle{ACM-Reference-Format}
\bibliography{refs}

\end{document}